\documentclass{PoS}
\usepackage{epsfig}

\def\hat{\widehat}
\def\chpt{\raise0.4ex\hbox{$\chi$}PT}
\def\schpt{S\raise0.4ex\hbox{$\chi$}PT}

\def\stag{{\rm stag}}
\def\cont{{\rm cont}}
\def\QCD{{\rm QCD}}

\def\MeV{{\rm Me\!V}}
\def\GeV{{\rm Ge\!V}}
\def\msbar{{\overline{\rm MS}}}

\newcommand\tr{{\rm tr}}

\def\spose#1{\hbox to 0pt{#1\hss}}
\def\ltapprox{\mathrel{\spose{\lower 3pt\hbox{$\mathchar"218$}}
 \raise 2.0pt\hbox{$\mathchar"13C$}}}
\def\gtapprox{\mathrel{\spose{\lower 3pt\hbox{$\mathchar"218$}}
 \raise 2.0pt\hbox{$\mathchar"13E$}}}
\def\inapprox{\mathrel{\spose{\lower 3pt\hbox{$\mathchar"218$}}
 \raise 2.0pt\hbox{$\mathchar"232$}}}

\title{Rooted staggered fermions: good, bad or ugly?}

\ShortTitle{Rooted staggered fermions: good, bad or ugly?}

\author{Stephen R.~Sharpe \\
  Department of Physics, University of Washington, Seattle,
  WA 98195-1560, USA \\
  E-mail: \email{sharpe@phys.washington.edu}}

\abstract{%
I give a status report on the validity of the 
so-called ``fourth-root trick'', i.e. the procedure of 
representing the determinant for a single fermion
by the fourth root of the staggered fermion determinant.
This has been used by the MILC collaboration 
to create a large ensemble of lattices using which
many quantities of physical interest have been and are being
calculated. It is also used extensively in studies
of QCD thermodynamics. The main question is whether the
theory so defined has the correct continuum limit.
There has been significant recent progress 
towards answering this question. 
After recalling the issue, and putting it into a
broader context of results from statistical mechanics,
I critically review the new work. I also address the
related issue of the impact of treating valence and
sea quarks differently in rooted simulations,
discuss whether rooted simulations at finite temperature
and density are subject to additional concerns,
and briefly update results for quark masses using
the MILC configurations.
An answer to the question in the title is proposed
in the summary.
}

\FullConference{XXIVth International Symposium on Lattice Field Theory\\

                July 23-28, 2006\\

                Tucson, Arizona, USA}

\begin{document}
\section{Introduction}
\label{sec:intro}

The field of lattice QCD is at a strange juncture in its history.
A significant fraction of the community is using an approach 
(``rooted staggered fermions''---to be explained in detail below)
which others in the community feel rests
on weak theoretical foundations, and yet others think may be wrong.
There is a lot at stake because simulations using rooted staggered
fermions have been computationally faster than those with other fermions
(although improvements in algorithms may have closed this gap)
and are already able to make accurate calculations of a number of
physical quantities. Results for ``gold plated'' masses and 
form factors (i.e. those which can be calculated most straightforwardly,
with systematic errors controlled) 
agree with experimental results within the claimed few percent 
accuracy~\cite{Davies,MILCmq,MILCfKpi,MILCsemil,MILCfD},
and several successful predictions have been made
(as summarized in Ref.~\cite{Kronfeld}).
Rooted staggered lattices are presently
being used for many ``mixed action'' calculations involving 
valence quarks with a different action, most notably domain-wall fermions
with an almost exact chiral symmetry.
Applications to light and heavy-quark flavor physics
and nucleon physics are described in other talks
at this meeting~\cite{Leelat06,Onogilat06,Orginoslat06}.

Thus, in the spirit of a well-known western, we need to know,
and the broader particle and nuclear physics communities need to know,
whether the results using rooted staggered fermions are
\begin{itemize}
\item {\bf GOOD}, i.e. have the correct continuum limit, 
without any complications;
\item {\bf BAD}, i.e. have the wrong continuum limit; or
\item {\bf UGLY}, i.e. have the correct continuum limit, but
with unphysical contributions present for $a\ne 0$
($a$ being the lattice spacing),
requiring theoretical understanding and complicated fits.
\end{itemize}
Let me stress that, 
if the continuum limit is wrong, then the results 
from rooted simulations are simply wrong, and there is
no {\em a priori} reason for them to be close
to those of QCD. In particular, it would make no sense to try and 
estimate systematic errors due to rooting. 
The situation differs from that with the quenched approximation 
(or the much better approximation of using $N_f=2$ light dynamical
flavors while quenching the strange quark) where one does expect many
results to be close to those of QCD. 
Of course, it is always possible that rooted simulations are ``bad''
but give results numerically close to those of QCD for some quantities---in other words,
that the numerical successes to date are a fluke.
It is because of this possibility, and because it is hard to make a
convincing argument by numerical results alone, that I focus
on theoretical arguments in this talk.
For more discussion of numerical tests see
last year's review by D\"urr~\cite{Durrlat05},
and the recent work of Refs.~\cite{Marescalat05,Hoffman06,Hart06,Numerical06}.

During the last year,
there has been significant theoretical progress 
towards understanding whether rooting is valid,
with issues clarified and in some cases resolved. 
There are, however, no quick and easy answers.
Many of the relevant papers are long and detailed,
and this appears to be unavoidable given the 
technical nature of the issue.
The result is that, although I have attempted to give only summaries,
and pick out the essential points,
this article is itself long and somewhat technical. 
Because of this, I provide a thorough overview
of the organization of the article in the following,
so as to allow a reader to skip to the core sections
if so desired.

I begin in the next section by collecting
the (uncontroversial) general assumptions that I make.
I then briefly recall the definition and general
properties of staggered fermions (sec.~\ref{sec:SF}),
with some details on perturbation theory relegated to app.~\ref{app:PT},
and explain what ``rooting'' means and why it is used
(sec.~\ref{sec:rootedSF}).
This brings me in sec.~\ref{sec:BGS}
to the first discussion of new work, 
namely the argument of Bernard, Golterman and Shamir
that rooted staggered fermions cannot be described by
a local single-taste theory except in the continuum limit~\cite{BGS}.
The main problem faced by rooting is now identified: non-locality.

At this point I switch gears, and give in sec.~\ref{sec:update} an update on
the status of the MILC ensemble of rooted staggered lattice configurations,
and the results obtained with them that are not discussed in other
plenary talks at this meeting.

Returning to my main theme, sec.~\ref{sec:SM} discusses what
we learn about non-local theories from studies of
lower-dimensional statistical mechanics systems.
The subsequent section, sec.~\ref{sec:tame}, is the core of this talk,
where I describe the arguments which have been given to understand and
``tame'' the non-locality. These use perturbation theory (PT),
renormalization group (RG) methods (the work of Shamir~\cite{ShamirRG06}), 
and chiral perturbation theory (\chpt) (the work of Bernard~\cite{CB06}).
I also discuss the numerical evidence concerning the $\beta-$function 
with rooted staggered fermions.

In sec.~\ref{sec:diseases} 
I describe some potentially serious problems 
with rooted staggered fermions which have been raised
by Creutz~\cite{Creutz06,Creutzlat06} 
and, I think, answered by Bernard, Golterman, Shamir and myself~\cite{BGSS}
and by D\"urr and Hoelbling~\cite{DH}.
The residue of these concerns is a clearer understanding of
the ugly features of rooted staggered fermions. 
A different concern (the ``valence rooting issue'')
is discussed in sec.~\ref{sec:valence}, with most of the
details of its resolution collected in app.~\ref{app:valence}.

Most of the work in the last year has argued for the
correctness of rooting. An important exception is the study
of non-zero density where Golterman, Shamir and Svetitsky
have observed that there is an additional problem~\cite{GSS}.
I discuss this in sec.~\ref{sec:GSS}, as well as the status
of rooting in simulations at non-zero temperature.

I summarize and give my conclusion---good, bad or ugly?---in sec.~\ref{sec:summ}.

A reader wishing to focus on developments on the central
issue can skip secs.~\ref{sec:assumptions},
\ref{sec:update}, \ref{sec:SM}, \ref{sec:diseases},
\ref{sec:valence} and \ref{sec:GSS}, as well as the appendices.
Someone familiar with staggered fermions
and the meaning of rooting can also
skip secs.~\ref{sec:SF} and \ref{sec:rootedSF}.
Put the other way, the key sections are 
\ref{sec:BGS}, \ref{sec:tame}, and the conclusions.

\section{Assumptions}
\label{sec:assumptions}

In this talk I will assume several results that,
while not proven, are generally accepted.
For the sake of clarity and completeness I collect them here.

The first assumption is the foundation of the field:
that lattice QCD with any standard choice of fermion
other than staggered
(i.e. Wilson-like [including clover and twisted mass], 
overlap and domain-wall)
has a universal, non-trivial continuum limit.
In other words, I assume that these ``uncontroversial'' fermion
discretizations are ``good'', in my classification above.
There is no rigorous proof of this, but it is strongly supported
by perturbation theory (asymptotic freedom plus
scaling of irrelevant operators) and extensive numerical,
non-perturbative tests.
The best that one could hope for in a discussion of rooted staggered
fermions would be to show that they have the same continuum limit
as that of one of the uncontroversial fermions.

The second assumption is that staggered fermions without
the rooting trick (which I will refer to by the ungainly
but descriptive term ``unrooted'') are uncontroversial,
and have a continuum limit with four degenerate fermions,
conventionally called ``tastes''.\footnote{%
Tastes were, in the past, referred to
as ``staggered flavors'' or simply ``flavors''.
``Tastes'' is preferred since, in the standard usage described below,
the four tastes are reduced to one (by rooting), and flavor
is introduced as an additional degree of freedom.}
The arguments for this are on a slightly weaker theoretical
footing than with Wilson-like fermions, but are nonetheless strong,
as I discuss in passing below.

My final working assumption will be that the
violations of reflection positivity that usually
accompany the use of improved fermion (or gauge) actions 
are unimportant for physical quantities.
In other words, while they give rise to unphysical 
degrees of freedom (e.g. with
complex energies or negative metric), these are restricted
to the cut-off scale.
I mention this point because essentially all staggered
simulations use improved actions (as do simulations with other 
fermion actions).

\section{What are staggered fermions?}
\label{sec:SF}

Euclidean staggered fermions have a very simple and economical action.
In unimproved form\footnote{%
In practice, as noted above, simulations use improved forms of
$D_\stag$, involving smeared gauge-fields and a three-step
(or knight's move) derivative. These improvements do not
impact the following considerations, so I do not show them
explicitly.}
this is~\cite{Susskind}
\begin{equation}
\bar\chi D_\stag \chi     = \sum_n \bar\chi_n 
\left[\sum_\mu \frac{\eta_{n,\mu}}{2}
\left( U_{n,\mu} \chi_{n+\mu} - U_{n-\mu,\mu}^\dagger \chi_{n-\mu} \right)
+ m_0 \chi_n \right]
\,,\label{eq:Sstag}
\end{equation}
where $n$ labels sites in a 4 dimensional lattice, $\chi_n$ are
{\em one-component} fermions, $U_{n,\mu}$ are the usual gauge
links and $m_0$ is the bare lattice mass (in lattice units,
so $m_0= a m_\cont$).
The Dirac matrices are cleverly packaged into the position
dependent signs $\eta_{n,\mu}$, whose form I will not need.

This is the action that is simulated,${}^2$ with numerical efficiency
resulting from the single component nature of the fermions
(as compared to the usual four Dirac components).
But this action does not make manifest the physical degrees
of freedom. It suffers from the doubling problem,
having $2^4=16$ one-component fermions in the classical continuum limit.
These can be packaged into 4 tastes of 4-component Dirac 
fermions.
One way to do this is to divide the lattice into $2^4$ hypercubes,
labeled by $n_y$ (with all components even), 
with the sites inside each hypercube labeled
by a ``hypercube vector'' $B$ with binary 
components~\cite{Gliozzi,Kluberg-Stern}.
In the free theory, the 16 $\chi$'s in each hypercube are
packaged into a Dirac-taste matrix as follows:
\begin{equation}
Q_{\beta,b}(n_y) = \frac18 \sum_B [\gamma_B]_{\beta,b}\ \chi_{n_y+B}\,;
\qquad
\gamma_B = \gamma_1^{B_1} \gamma_2^{B_2} \gamma_3^{B_3} \gamma_4^{B_4}
\,.
\label{eq:Qdef}
\end{equation}
The free staggered action [eq.~(\ref{eq:Sstag}) with $U_{n,\mu}=1$] can
be exactly rewritten as follows (using momentum space for later convenience):
\begin{equation}
\sum_{p_y} \overline Q(p_y) \left\{
\underbrace{\left[ \sum_\mu i 
(\gamma_\mu\otimes {\bf 1} )\sin p_{y,\mu}
\!+\!({\bf 1}\otimes {\bf 1}) m_0 \right]}_{{ O(a)}}
\!+\! 
\underbrace{\sum_\mu (\gamma_5 \otimes \xi_\mu \xi_5) 
(1 \!-\! \cos p_{y,\mu})}_{{ O(a^2)}}
 \right\} Q(p_y)
\,.
\label{eq:SstagQ}
\end{equation}
Here $p_y$ is the momentum (in lattice units)
 conjugate to the hypercube vector $n_y$,
and so does not include oscillations {\em within} the hypercubes.
Such high momenta correspond to the doubler modes on the original lattice
and are here included as degrees of freedom in $Q$ itself.
The notation $(\gamma_B\otimes \xi_C)$, which will recur repeatedly in this talk,
indicates the matrices acting on the Dirac ($\gamma_B$) and taste ($\xi_B$) indices of
$Q_{\beta,b}$, namely $\beta$ and $b$ respectively.
The taste generators $\xi_B$ are defined like $\gamma_B$ in (\ref{eq:Qdef}),
but with $\xi_\mu=\gamma_\mu^\star$.
(The complex conjugation here is a convenient, but not
fundamental, convention.) The $\xi_B$ form a basis for the generators of 
the taste group $SU(4)$.

In the form (\ref{eq:SstagQ}), the (free) staggered action looks like 
four Wilson fermions with an unconventional Wilson term.
If the Dirac-taste matrix in the Wilson term were replaced
by $({\bf 1}\otimes {\bf 1})$ then one would simply have four Wilson fermions.
With the form shown, however,
the Wilson term both removes the doubler modes 
(those with $p_{y,\mu}\approx \pi$) from the continuum spectrum
{\em and} breaks most of the taste symmetries (both vector and axial).
What is gained by breaking these taste symmetries is that one 
{\em axial} generator
remains unbroken by the Wilson term, that which generates the
so-called $U(1)_\epsilon$ symmetry:
\begin{equation}
Q \to \exp[i\alpha (\gamma_5\otimes\xi_5)]\,,\qquad
\overline Q \to \overline Q \exp[i\alpha (\gamma_5\otimes\xi_5)]\,.
\end{equation}
This symmetry remains in the interacting theory, where it must be
defined at the level of the $\chi$ fields.
The situation should be contrasted with
standard Wilson fermions, where the Wilson term
breaks all the axial but none of the vector symmetries.
The importance of the $U(1)_\epsilon$ symmetry is that, unlike with
standard Wilson fermions, there is no renormalization of the critical mass---
it remains at $m_0=0$. This is one of the practical advantages of staggered fermions.

As indicated by the underbraces in eq.~(\ref{eq:SstagQ}),
the Wilson term is irrelevant in the continuum limit when $p\sim O(a)$.
One's naive expectation is thus 
that the $SU(4)$ taste symmetry is restored as $a\to 0$,
just as the broken axial symmetries are restored with normal Wilson fermions.
This argument is naive because, in the interacting theory, the packaging
transformation (\ref{eq:Qdef}) must include gauge fields to retain gauge
invariance, and the gauge fields living ``inside'' the $Q$'s can also lead to
taste breaking. A careful
analysis must enumerate possible relevant operators consistent
with the symmetries of the {\em original} lattice action. 
Such an analysis~\cite{GS}
finds no additional relevant operators, which, 
coupled with asymptotic freedom,
implies that the symmetry restoration in the free theory applies also 
in the presence of interactions. 
This is the basis for my assumption that ``unrooted'' staggered
fermions have the expected continuum limit. 

The description so far uses what is called, for obvious reasons,
the ``position-space basis''---spin and taste are distributed over
a space-time hypercube. For many applications, and in particular for
perturbative calculations, it is more convenient to use a different basis,
one set up in momentum space. This is described in app.~\ref{app:PT}.

\section{Rooted staggered fermions}
\label{sec:rootedSF}

We do not want a continuum theory with four degenerate quarks---we want
QCD with non-degenerate quarks. 
The obvious approach is to somehow make the tastes non-degenerate
in the continuum limit, so that they could correspond to the $u$, $d$,
$s$ and $c$ quarks. 
This can be done by adding additional 
terms to the action (involving $\chi$ and $\bar\chi$ on different,
but nearby, sites)~\cite{BJ,Mitra,GS}.
Furthermore, the coefficients of each of these terms is
multiplicatively renormalized~\cite{GSold,Gockeler,MitraWeisz,GS},
i.e. the lattice symmetries remain strong enough to forbid
mixing with the identity operator.
This approach has not, however, been followed in practice,
and it is perhaps worthwhile discussing why.

I think this is for four main reasons.
First, there is a tuning problem. The physical masses
are linear combinations of the coefficients which are 
multiplicatively renormalized. Since we only know these
renormalizations approximately (e.g. in one-loop perturbation theory),
the coefficients would have to be tuned numerically to obtain the
correct physical masses.
Perhaps more important is that, although one can pick mass terms
such that the determinant is real, it is not necessarily
positive, and its spectrum has a general complex form.
This is compared to the real, positive determinant of staggered
fermions with its spectrum lying on a line
(parallel to the imaginary axis) and bounded away from the origin.
Thus at least part of
the numerical advantage of staggered fermions would likely be lost.
Third, since the general masses break some of
the lattice symmetries, continuum multiplets 
fragment into even smaller lattice representations.
And, finally, the action would be complicated to
simulate.

What is done in practice is to use 
one staggered fermion for each flavor and to
take the fourth-root of the fermion determinant:\footnote{%
This method was introduced to study the Schwinger model
in Ref.~\cite{MPR}.}
\begin{equation}
Z_\QCD^{{\rm root}} = \int DU e^{-S_{\rm gauge}}
\left\{\det[D_\stag(m_u)] \det[D_\stag(m_d)] 
\det[D_\stag(m_s)]\right\}^\mathbf{ 1/4}
\,,
\label{eq:ZQCDroot}
\end{equation}
where I have restricted myself to the three light flavors.\footnote{%
Most simulations to date work in the isospin symmetric limit 
with $m_u=m_d$,
so the two fourth-roots combine to a square-root in the up-down sector.}
Note that $\det(D_\stag)$ is positive definite for any non-zero quark mass
(eigenvalues come in pairs due to the $U(1)_\epsilon$ symmetry:
$\pm i \lambda + m_0$ with $\lambda$ real), 
and one always takes the positive fourth-root.

The rationale for ``rooting'' is that, in the continuum limit,
each $D_\stag$ represents four degenerate tastes, and thus has
the structure $D_1\otimes {\bf 1}$, where $D_1$ is a one taste operator.
If so, $\det[D_\stag]={\det}^4(D_1)$, and rooting is legitimate.
For $a\ne 0$, however, taste is broken and this argument fails.
We thus arrive at the central question addressed in this talk:
\vspace{-.2cm}
\begin{center}
\fbox{\parbox{4truein}{
{\bf Is it legitimate to take the fourth-root of the fermion determinant
before sending $\mathbf{a\to0}$?}
}}
\end{center}

\section{Non-locality of single-taste action and its implications}
\label{sec:BGS}

The basic problem with rooting has been clear from the beginning: locality.
The determinant of a local operator can be written
as a fermionic functional integral with a local action---this is how the
determinant arises in the first place.
Locality is one of the properties required of the action to
ensure that the underlying theory is physical at long distances.\footnote{%
See the review by Jansen~\cite{Jansenlat03} 
for a thorough discussion of all the properties that
are needed and their status with different fermion discretizations.}
The fourth-root of the determinant of a general local operator 
without an $SU(4)$ symmetry cannot, however,
be written in any obvious way as a functional integral over a fermion with
a local action. It can certainly be written as a functional integral,
e.g. 
with the action $\bar\chi_{\rm bad} (D_\stag)^{1/4} \chi_{\rm bad}$
in the present case, 
but the action here is non-local (even in the free theory)~\cite{Bunk}.
The correct question is whether there is a clever way of manipulating $D_\stag$ to
arrive at a local fermionic formulation. 
The answer is now clear from the work of Bernard, Golterman
and Shamir (BGS)~\cite{BGS}: {\bf NO}. In particular,
\vspace{-.2cm}
\begin{center}
\fbox{\parbox{4truein}{
{\bf Rooted staggered fermions cannot be described by a local theory
with a single taste per flavor.}
}}
\end{center}

The argument for this is straightforward, and proceeds by contradiction.
The most general way to represent the rooted determinant by
a local single-taste theory requires
\begin{equation}
\left(\det[D_\stag]\right)^{1/4} = \det[D_1] \exp(- \delta S_{\rm eff,g})
\,,
\label{eq:Adams}
\end{equation}
where $D_1$  a local, single-taste operator and $\delta S_{\rm eff,g}$ 
is a local gauge action. The latter is allowed 
because $\delta S_{\rm eff,g}$ can be absorbed into the gauge 
action~\cite{Adams,ShamirRG04}. 
The result (\ref{eq:Adams}) is assumed to hold on all configurations,
a condition which might be relaxed, as discussed below.
It follows that the fermionic weight for ``unrooted''
staggered fermions is
\begin{equation}
\det[D_\stag] = \left(\det[D_1]\right)^4 \exp(- 4\;\delta S_{\rm eff,g})
= \det(D_1 \otimes \mathbf{1}) \exp(- 4\;\delta S_{\rm eff,g})
\label{eq:BGS}
\,.
\end{equation}
In words, this result implies that the unrooted 
staggered theory can be rewritten
as a local theory with an exact $SU(4)$ taste symmetry.
But we know that this is false. 
The staggered action retains only a discrete subgroup
of the $SU(4)$ taste symmetry, 
and thus its spectrum will not fall into $SU(4)$ multiplets.
In particular, while the lightest states in
the $SU(4)$ symmetric theory will be
a 15-plet of pseudo-Goldstone bosons (PGBs), the staggered
theory has only a single PGB (associated with the $U(1)_\epsilon$),
with the other 14 
light pseudoscalars falling into multiplets of the lattice
symmetry with masses differing from the PGB by $O(a^2)$.
For an example of a concrete contradiction consider the 
connected part of the two-point correlator of the plaquette.
At long distances this will be dominated by
two-pseudoscalar contributions, and there is clearly no way 
that these can match over a range of distances
given the different masses of the states in the two theories.

Thus we learn that the starting assumption, eq.~(\ref{eq:Adams}), is false:
a local one-taste formulation is not possible. 
Any one-taste formulation of rooted staggered fermions must be non-local.
A non-local action can lead to a resolution 
of the contradiction because the notion of
a spectrum (needed in the argument above) can no longer be used.
An explicit example of such a resolution will
be seen in sec.~\ref{sec:RG}
[see eq.~(\ref{eq:rooteddetcont}) and subsequent discussion].

An obvious corollary of the BGS result is that,
if rooted staggered fermions are to yield a physical
continuum limit describing a single taste,
the non-locality must vanish when $a\to0$.
How this can occur is the focus of sec.~\ref{sec:tame} below.
I only note here that it is consistent with
eq.~(\ref{eq:BGS}), because the spectrum on
the l.h.s. does become $SU(4)$ symmetric 
in the continuum limit (given the assumption that
the unrooted staggered theory is ``good''), 
and so a local $\delta S_{\rm eff,g}$ suffices.


The BGS argument is not completely watertight. The relation (\ref{eq:Adams})
needs to hold only on the subset of configurations that are important
in the rooted ensemble, and it could be that this subset has no overlap
with that which is important in the unrooted ensemble. 
This would invalidate the necessity of non-locality.
I am not sure how plausible such a loophole is, but in any case 
I do not pursue this possibility because the
RG and \chpt\ arguments in sec.~\ref{sec:tame} imply that rooted
staggered fermions are indeed non-local for $a\ne0$, 
consistent with the conclusion of the BGS argument.

\bigskip

Rooted staggered fermions have reached their nadir at this point in the talk.
A reasonable person might say:
\begin{quote}
\emph{``Locality of the action guarantees universality, i.e. that we obtain
the correct continuum limit. Non-local theories are unphysical
(lacking unitarity, \dots). The possibility that the non-locality vanishes
in the continuum limit sounds hard to establish, even implausible.
I don't want to use rooted staggered fermions.''}
\end{quote}
This is not, I think, the only reasonable reaction. A second reasonable person
might say:
\begin{quote}
\emph{``Rooted staggered fermions are so attractive numerically, that I am
going to try and understand and ``tame'' the non-locality. If I can argue
plausibly that the non-locality does not change the universality class,
i.e. that all the effects of non-locality vanish in the continuum limit,
then the extensive numerical results based on the MILC ensemble
will be physical, as long as the chiral and continuum extrapolations
take into account the effects of non-locality.''}
\end{quote}
Most of the rest of the talk will concern the attempts to
tame the non-locality. First, however, I want to recall why the
stakes are so high, and at the same time briefly change roles into
a reviewer of the status of staggered simulations.\footnote{%
For an update on the status with 
other fermions see the talk by Giusti~\cite{Giustilat06}.}

\section{Update on rooted staggered simulations}
\label{sec:update}

In this section I will set aside the theoretical questions about rooting,
and review numerical progress with rooted staggered simulations.

Over the last 5 or so years,
the MILC collaboration has created a large ensemble of configurations
using the rooting trick with a particular variant of improved staggered fermions
(the ``asqtad'' action---$a^2$ terms removed at tree-level
from both fermion and gauge actions, and tadpole improved~\cite{asqtad}).
Aside from the concerns about rooting,
this ensemble has been very successful and is a model for future ensembles
using other fermion discretizations. The most important features are:
\begin{enumerate}
\item
All three light quarks are unquenched, i.e.
these are fully unquenched simulations.
In practice, ``$2+1$'' flavors are used, with $m_\ell=m_u=m_d$,
so that isospin is exact.
This is a good enough approximation
for few percent accuracy in many physical quantities.
\item
The light quarks are truly light, ranging down to
$m_\ell \approx m_s^{\rm phys}/10$. Having good statistics at
such low masses is crucial for accurate chiral extrapolations.
\item
There are large numbers (of order 500) of independent
lattice configurations for each value of the parameters,
so that statistical errors in many quantities are smaller than
systematic errors.
\item
The lattices are widely available, so that many quantities
of physical interest can and have been calculated on the ensemble
by many collaborations. 
This feature of the MILC ensemble
has set a precedent that is happily being followed
quite widely, facilitated by the ILDG (International Lattice Data Grid),
as reviewed here by Jansen~\cite{ILDGlat06}.
\end{enumerate}
Until recently, one of the weakness of
the MILC ensemble was that it lacked sufficient lattice
spacings to allow a thorough test of the continuum extrapolation
(particularly important in light of the non-locality issue).
Previously there were ``fine'' lattices with $a\approx 0.09\;$fm
and ``coarse'' lattices with $a\approx 0.12\;$fm. Now there are
also ``medium coarse'' ($a\approx 0.15\;$fm) 
and the beginnings of a ``super-fine'' ($a\approx 0.06\;$fm) set. The latter
is clearly the most important and will be filled out substantially this year.
There are also ``very coarse''
($a\approx 0.18\;$fm) lattices, but for many quantities
the taste-breaking may be too large for these lattices to be useful.
A complete list of available lattices is available in the contribution
of Sugar~\cite{Sugarlat06}.

What remains as a shortcoming is that only in one case is it possible
to make a comparison between two different volumes. This is not a criticism of
the MILC collaboration---they have simply
had to make hard choices given the large computational resources required.
The problem is well illustrated by the size of the super-fine lattices:
$48^3\times 144$ and up!

\begin{figure}
\centerline{\epsfxsize=5.6in\epsfbox{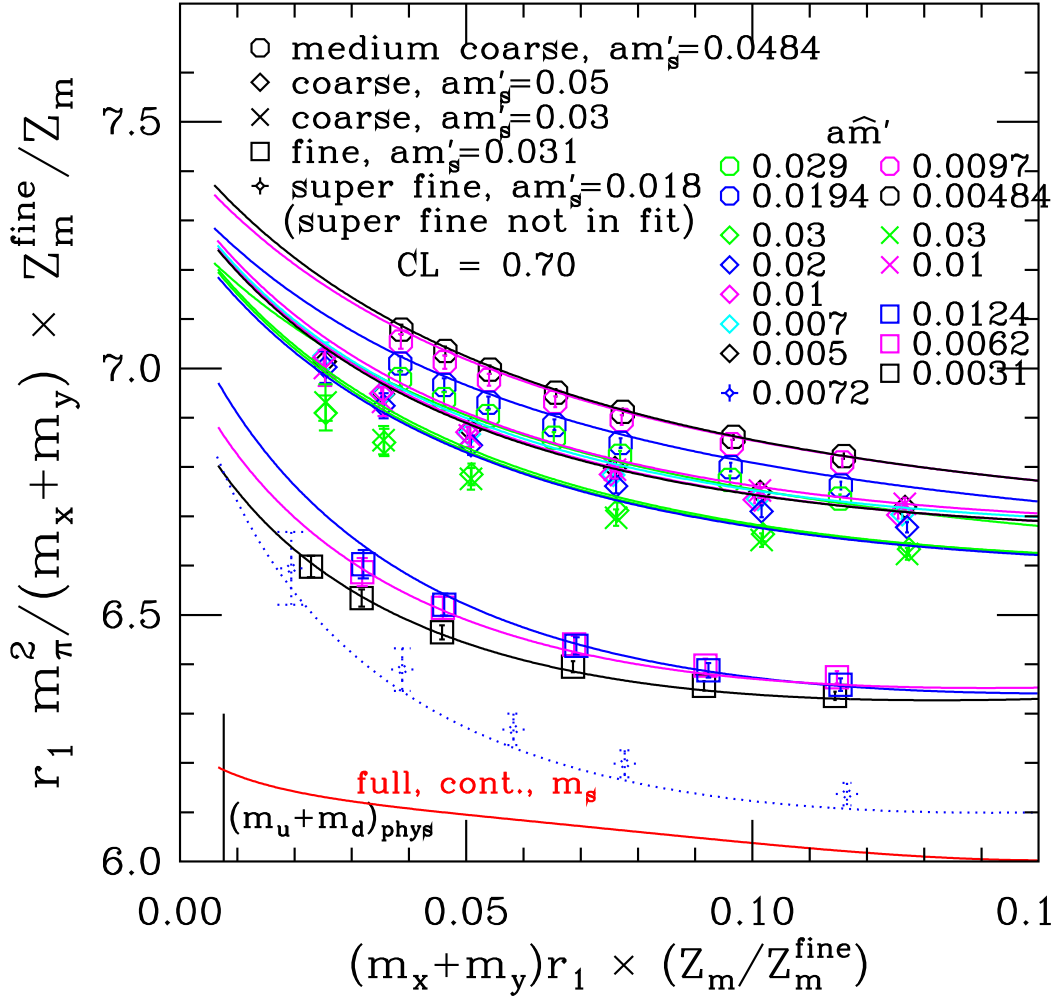}}
\caption{Update on pion mass results using rooted improved staggered
fermions. The scale is set with $r_1$. $Z$-factors account for
logarithmic scaling of quark masses. See text for further explanation.}
\label{fig:MILC}
\end{figure}

As noted above, these lattices are being used for a wide range of
physics applications.
These are discussed in the plenary talks on flavor
physics~\cite{Leelat06,Onogilat06} and hadronic physics~\cite{Orginoslat06}.
It is clearly crucial, however, that the systematics of taste-breaking
are numerically well understood in simple quantities if we are to
trust the results for more complicated applications.
To illustrate the status I show in Fig.~\ref{fig:MILC} results
for the ratio $m_\pi^2/(m_x+m_y)$ versus the sum of the
valence quark masses $m_x+m_y$~\cite{Sugarlat06}.
Almost all the points are partially quenched (PQ), with the upper
cluster (circles) being on the medium-coarse lattices,
the middle cluster (diamonds and crosses) from the coarse
lattices,  and the lower cluster (squares) from the fine lattices.
The lines through the points are part of a global fit to $m_\pi$, $m_K$, $f_\pi$
and $f_K$, a generalization of 
that described in Ref.~\cite{MILCfKpi}.
The fits are to (rooted) staggered \chpt,
a theory discussed below, and have good confidence levels.
The predicted extrapolation to the continuum limit, including an
interpolation to the physical strange quark mass, is shown as the 
(red) line at the bottom. The vertical line at about 0.008 on the horizontal axis
shows the resulting physical average light quark mass.
I stress that the continuum limit result shown in Fig.~\ref{fig:MILC} is
not PQ---it is for a pion composed of degenerate sea quarks of mass $m_x=m_y=m_\ell$.

The main point I want to make with this plot is that the fit
allows one to predict where the data from the single super-fine
lattice should lie. This prediction is shown by the dotted (blue) line, 
along with the new data (dotted fancy pluses).
Note that (in contrast to the continuum prediction discussed above)
this is a prediction for a PQ pion (sea quark masses held fixed and
valence masses $m_x$ and $m_y$ varying).
The predicted curvature is an example of
an enhanced PQ logarithm~\cite{SSenhancedlog}.
The prediction works well (remembering that all these PQ points
are correlated). 

I draw two conclusions, one staggered-specific and
one general. First, PGB properties from rooted staggered simulations can
be well described using staggered \chpt. This 
numerical test could have failed, as illustrated by complete failure of
fits to continuum \chpt~\cite{MILCfKpi}.
Second, this is one example of a growing body of evidence
that PQ\chpt\ can successfully describe PQ simulations
(irrespective of the type of fermions). This observation will
be useful below.

The one physics result that I want to quote is an update
on values for quark masses:
\begin{eqnarray}
m_s^\msbar(2\;\GeV) = 90(0)(5)(4)(0)\;\MeV &&  [87(0)(4)(4)(0)\;\MeV]\\
m_s/\hat m = 27.2(0)(4)(0)(0) &&  [{27.4(1)(4)(0)(1)}] \\
m_u/m_d = 0.42(0)(1)(0)(4)  && [{0.43(0)(1)(0)(8)}]
\,.
\end{eqnarray}
The new results (still preliminary)
include the super-fine and medium-coarse lattices~\cite{Sugarlat06},
and should be compared to the old results [shown in parentheses]
which used only the coarse and fine lattices~\cite{MILCmq,MILCfKpi,Masonmq}.
The four errors are from (i) statistics, (ii)
fitting and other systematics of the simulations, (iii)
perturbation theory, and (iv) the uncertainty in the estimates of EM effects.
The point I want to emphasize here
is that the inclusion of the first super-fine lattice 
changes the results at a level consistent with the quoted errors.
It is clearly important to see whether this remains true
as the super-fine ensemble is extended.

I choose this result as it is an example of a (potentially) successful
completion of one of the key goals of lattice simulations:
determination of the fundamental parameters of QCD.
Knowing $m_s$ to better than 10\%, and ruling out $m_u=0$
at the 10-$\sigma$ level, are impressive results.
But are they correct? One issue is the theoretical concern about
rooting---the focus of the rest of my talk. Aside from this,
there is the more mundane question of whether the two-loop
perturbative matching~\cite{Masonmq} 
that is employed is sufficiently accurate. 
The inclusion of the two-loop contributions
raised $m_s$ by 14\% compared to using one-loop matching,
and the estimated three-loop error 
($\pm 2 \alpha_s^3 \approx \pm 4\;$MeV in $m_s$)
is a guess, albeit a reasonable one.
It is clearly important to check the matching factor
using non-perturbative renormalization---calculations are underway,
and we should know by next year.

To date, there is no fully unquenched result
with ``uncontroversial'' fermions to which this 
rooted staggered result can be compared.
Nevertheless, there are interesting results with
$N_f=2$ flavors of improved Wilson fermions
that suggest important lessons.
In particular, they indicate that using
non-perturbative renormalization can 
change the result by an amount larger than the naive
guess of the size of higher-order perturbative corrections.
For example, Ref.~\cite{twoflavorms} finds that, at a lattice
spacing corresponding to the fine MILC lattices,
$m_s$ is 24\% higher with the non-perturbative Z factor 
than it would be with the one-loop value.

A second lesson concerns the lattice spacing dependence.
The simulations of Ref.~\cite{twoflavorms} 
have pushed to finer lattices than MILC,
and find a significant dependence on $a^2$ 
between the fine MILC spacing and the continuum limit. 
While there is no particular reason that the same should be true
with (rooted) staggered fermions, this underlines the
importance of completing the super-fine ensemble and possibly
pushing to yet finer lattices. 


As is well known, rooted staggered fermions have been used
to calculate a number
of ``gold-plated'' quantities, finding agreement with experiment
at the few percent level~\cite{Davies}. Furthermore, predictions
for the $D\to K\ell\nu$ form factor, $f_D$ and the $B_c$ mass
were subsequently confirmed by experiment 
(as summarized in Ref.~\cite{Kronfeld}).
While these successes are impressive, 
numerical checks cannot be definitive.
For one thing, the required fitting is complicated, as illustrated 
by Fig.~\ref{fig:MILC}.
We would want calculations with other fermion discretizations
even without the rooting issue. 
It could also be that the
agreement is a fluke---a wrong theory could give results close
to experiment for some quantities but not in general.
But the most important reason that numerical tests do not suffice
is that there is a serious theoretical problem (non-locality), 
and it must be understood.

\section{Non-local interactions in statistical mechanics}
\label{sec:SM}

I now take a brief excursion into statistical mechanics (SM) to
see what lessons can be learned about non-local interactions.
In particular, while for local interactions we expect
the universality class to be determined by the number of dimensions
and by the symmetries of the order parameter, this will not
be true in general with non-local interactions.

There is a considerable literature studying power-law interactions
in two and three dimensions, motivated in part by physical
interactions such as the van de Waals force. 
One well-studied example is the Ising-like model
in $d$-dimensions with Hamiltonian:
\begin{equation}
H = \mathcal{J} \sum_{\vec x,\vec y} 
s_{\vec x}\textrm{``} \frac{1}{|x-y|^{d+\sigma}}\textrm{''}  s_{\vec y}
\,.
\label{eq:Ising}
\end{equation}
The quotes around the power-law indicate that a smoothed non-singular
form is used, which has no impact on the long distance behavior.
${\cal J}$ is the coupling, and $\sigma$, a positive real number,
characterizes the non-locality. The non-locality
increases as $\sigma$ is reduced.

Such theories have been studied using all
the tools of SM and I quote some examples.
The original works used renormalization group (RG)
methods and the $\epsilon-$expansion~\cite{Fisher72,Suzuki72},
with recent extensions studying corrections to scaling,
e.g. Ref.~\cite{Dantchev01}.
There are a number of exact results, e.g. Ref.~\cite{Aizenman88}.
There have also been numerical simulations,
e.g. Refs.~\cite{Luijten97,Luijten02}, 
which, as can be imagined, are challenging for non-local interactions.
I do not have time or competence to discuss all this work, and
note only one simple point. In the RG analysis, after
blocking, the spins are replaced by scalar fields, which have
a $\phi^4$ interaction, and a bare propagator whose inverse at
long distances has the form:
\begin{equation}
G_0(k)^{-1} \sim r + j_\sigma k^\sigma + j_2 k^2 + \dots
\,,
\end{equation}
with $r$, $j_\sigma$ and $j_2$ calculable constants.
This is obtained by Fourier transforming the interaction
in (\ref{eq:Ising}), and holds as written for non-integer $\sigma$
(otherwise there are logarithmic corrections).
This expression shows that in the free theory, when $r$ is tuned
to its critical value of zero (by varying the temperature), 
the IR behavior is dominated by the non-local part of the 
interaction if $\sigma< 2$,
but, conversely, the non-local part is sub-dominant if $\sigma >2$.

\begin{figure}
\centerline{\epsfxsize=6in\epsfbox{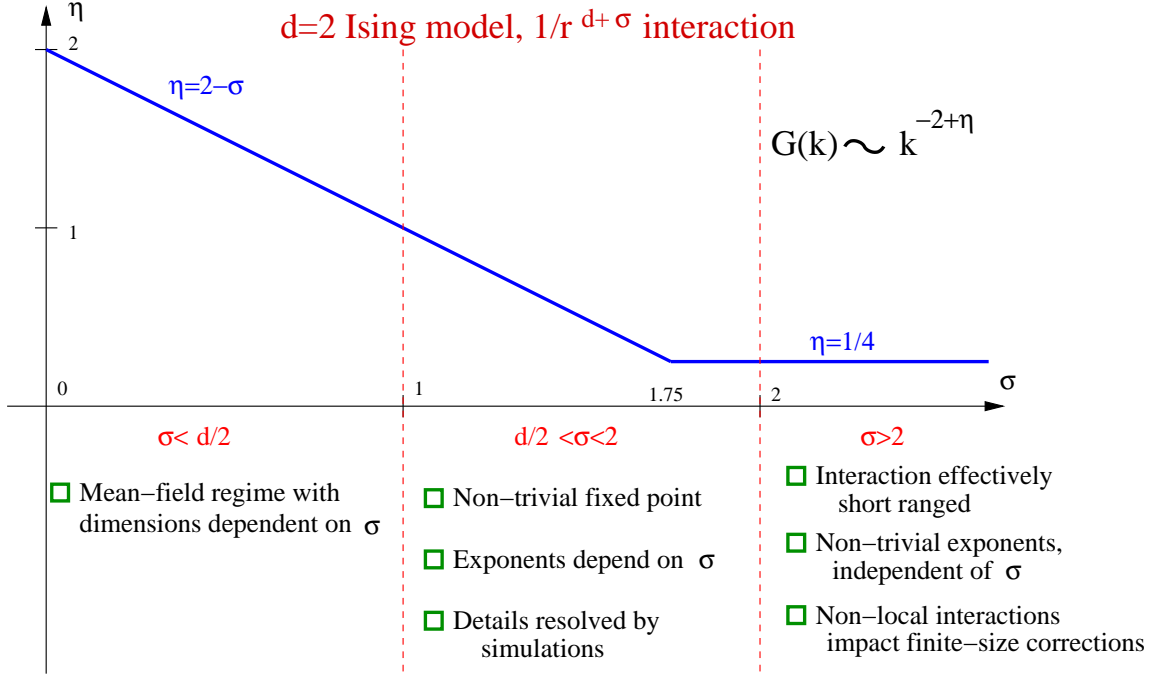}}
\caption{Results for correlation-length exponent for $d=2$ Ising model
with long-range interactions.}
\label{fig:Ising}
\end{figure}

These observations help one understand the results that
have been obtained for a variety of models with non-local interactions.
I have chosen the $d=2$ Ising-like model as a concrete example and
collect in Fig.~\ref{fig:Ising} results for the critical exponent $\eta$
(taken from Ref.~\cite{Luijten02} and references therein). 
Recall that $\eta$ is defined by 
the small $k$ dependence of the {\em full} propagator
at the critical point: $G(k) \sim k^{-2+\eta}$.
What one learns from an RG analysis is that, as suggested above,
for $\sigma>2$ the interaction falls off fast enough
that it is effectively short ranged, and one
finds the usual local critical exponents (here $\eta=1/4$).
The key point is that these are {\em independent of $\sigma$}.
It should be noted, however, that the non-local interactions
do impact finite-volume corrections. For example, in the disordered
phase above the critical temperature, and assuming periodic boundary
conditions,
finite-volume corrections are exponentially suppressed with local
interactions (as are corrections to hadron masses in QCD),
but are power-law suppressed, as $L^{-(d+\sigma)}$, 
with non-local interactions~\cite{Dantchev01}.

For $\sigma<2$, i.e. more slowly falling interactions,
the $k^\sigma$ term in the free propagator
dominates in the IR, changing the scaling analysis.
If $\sigma < d/2$ one finds that the $\phi^4$ interaction is
irrelevant, leading to mean-field-like behavior, with $G(k)\sim k^{-\sigma}$
at criticality.
Thus one finds $\eta=2-\sigma$, 
so that the critical exponent {\em does
depend on $\sigma$}. Crudely speaking, a non-local interaction
increases the effective number of neighbors, which is the same effect
as occurs when one increases the dimension,
and one is effectively pushed above the upper critical dimension 
(which is 4 for local interactions) into the mean-field regime.
Finally, for $d/2 < \sigma < 2$ the $\phi^4$ interaction is relevant
and the critical behavior non-trivial. The exponents can be estimated
using the $\epsilon$-expansion, and calculated using simulations.
They depend, in general, on $\sigma$, as in the example shown in the 
Fig.~\ref{fig:Ising}.

Clearly one must be careful in drawing lessons from these results.
For one thing, rooted staggered QCD has a dominant interaction,
that from the gauge sector, which is local.
For another, we are at the upper critical dimension, and so
corrections to scaling are logarithmic rather than power law
(which corresponds to the middle regime,
$d/2 < \sigma < 2$, being squeezed to a point).
Nevertheless, I draw the following four general lessons:
\begin{itemize}
\item
Non-local interactions need not change the universality class
(the $\sigma>2$ case above).
Thus there is no {\em a priori} reason to rule out rooted staggered
fermions. Instead, one must study the issue.
\item
Perturbation theory (which is the analog
of the RG equations in the SM examples)
is a useful tool for such a study.
\item
One should check the perturbative conclusions using simulations.
A crucial question is whether the scaling is as in the
desired continuum limit, i.e. whether $\alpha_s$ runs correctly.
\item
One should beware of enhanced finite size effects---standard \chpt\ 
predictions for such effects will likely fail. As we will see,
this can be remedied by using ``rooted'' staggered \chpt.
\end{itemize}
While we do not really need the SM example to come to these conclusions,
I find that it helps ``demystify'' the issue of non-locality.

\section{Can we tame the non-locality?}
\label{sec:tame}

I now come to the core of this talk.
We know that QCD with rooted staggered fermions involves non-local
interactions, and the key question 
is whether this non-locality changes the continuum limit.
I will discuss in turn what has been learned using perturbation theory,
numerical simulations, a general RG analysis, and \chpt.

\subsection{Perturbation theory}
\label{sec:PT}

In this section I discuss what
we learn about QCD with rooted staggered fermions
using perturbation theory.
I go into some detail because, although some of the
discussion has been given in the literature,
I am unaware of a complete account. 

I start from the assumed renormalizability of 
lattice QCD with unrooted staggered fermions.
That is, by adding the usual (finite number of) counterterms,
which depend logarithmically on $a$, 
one finds that properly normalized correlation functions 
with non-vanishing Euclidean momenta 
have a well-defined limit as $a\to 0$.
This limit is a power series in the renormalized coupling constant
(evaluated at a scale set by the external momenta).
The counterterms can be chosen to respect BRST symmetry.
Furthermore, their finite parts can be chosen so that
the resulting perturbative expansions are equal to
those for a theory with 4 degenerate
fermions in any desired scheme (e.g. the regularization
independent (RI) or $\overline{\rm MS}$ schemes).
This is all a long-winded way of saying that 
renormalization with lattice regularization works 
as with other perturbative regulators---
the additional vertices present on the lattice
($q\bar qgg$ etc.), and the taste violations that
are present at large gluon momenta, both
lead only to finite renormalizations with a taste-symmetric form.

Renormalizability extends straightforwardly to a theory with $n_r$
copies, or ``replicas'', of unrooted staggered fermions,
with $n_r$ a positive integer.
Counterterms corresponding to divergent fermion loop contributions
are simply multiplied by $n_r^{{\# \rm loops}}$, and one obtains
renormalized correlators for a theory with $4 n_r$ continuum fermions.
Thus far the discussion is uncontroversial.
The next step is a little less so: I will assume
that the ``fermion-loop'' rule can be applied unambiguously not only to 
the divergent parts, but also to the finite parts of counterterms
needed to bring renormalized correlators into the desired 
scheme.\footnote{%
I am excluding here hypothetical schemes in which the
renormalization conditions have an explicit non-polynomial
dependence on $n_r$.}
This assumption is important, 
because it implies that, at any order in PT,
the counterterms needed to arrive at the desired 
scheme are polynomials in $n_r$.\footnote{%
To be completely clear, the issue is not whether 
counterterms can be chosen so that the correlators
are in the desired scheme---this is a corollary of
the assumed renormalizability---but rather whether the
desired counterterms are polynomial in $n_r$.}
This claim strikes some (myself included)
as very reasonable and others as less obvious. 
What seems clear, however, is that a demonstration (or refutation)
of this supposition would be greatly facilitated by having
a proof of the renormalizability of unrooted staggered fermions in hand.

The importance of this discussion is that it allows one to
extend the considerations to non-integer $n_r$, as first noted
in Ref.~\cite{BGPQ}.
The argument goes as follows.
Imagine doing a simulation with fermionic weight factor ${\det}^{n_r}(D_\stag)$,
which I will generically call the ``rooted theory''.
In perturbation theory, for gluonic correlation functions, this has the
effect of multiplying each fermion loop by $n_r$ (compared to the unrooted case),
{\em irrespective of whether $n_r$ is an integer}.
Such correlation functions can be renormalized simply by substituting
the actual value of $n_r$ into the counterterms determined for integral $n_r$.
This follows because
both the bare correlators and the counterterms
are polynomials in $n_r$, and thus so are the renormalized correlators. 
Since, by assumption, the latter are finite for all positive integer $n_r$, 
the coefficients of the polynomial are finite, and thus
finiteness holds for all $n_r$. Analytic continuation from the positive
integers is possible because we know that the form is polynomial.

This establishes the renormalizability of gluonic correlators
in the rooted theory. Note that no additional types of counterterm are needed.
However, we need a stronger result, namely that the 
renormalized correlators are not only finite but are in fact
equal to those in the desired scheme with $N_f=4 n_r$ fermions for any $n_r$.
This follows from the assumption that the finite parts of
the counterterms are also polynomial in $n_r$.
It also uses the fact that 
correlators in the desired scheme with $N_f$ flavors are,
at any finite order in PT, polynomials in $N_f$. Since we know
that the lattice regularization reproduces these polynomials when $N_f=4 n_r$
with $n_r$ any positive integer, it also does so for any $n_r$.

There is one loose end in this argument---what about fermionic correlators?
Fermionic quantities in actual simulations are composed
of ``unrooted'' propagators, $D_\stag^{-1}$. Rooting removes the
extra degrees of freedom in the sea, but what about in the valence sector?
It seems that we are stuck with $4 n_r$ valence tastes, with integer $n_r$.
Does this not lead to violations of unitarity?
This is what I call the ``valence rooting'' issue
(which sounds better than the more descriptive ``valence unrooting'' issue).
I discuss it here in the context of PT and then
more generally in sec.~\ref{sec:valence}. 

To describe the extra valence quarks one is forced 
into a partially quenched framework.
If $n_r=1/4$, for example,
the theory would have a single sea taste but $4$ valence tastes.
In practice, one simply uses the full staggered propagators for valence
lines and multiplies loops by $n_r$.
The key point is that, by an extension of the previous argument, it
is very plausible that valence fermion correlation functions can
be renormalized and related to those in a
corresponding {\em continuum} PQ theory.
I claim further that taste will be unbroken in this continuum theory,
essentially because this is what happens in the unrooted staggered
theory (where the valence fields are, by definition, unrooted).
But in a PQ theory in which sea and valence quarks have the same
Dirac operator and masses (because taste is unbroken), there is
a unitary, physical sub-sector obtained by projecting onto
the appropriate number of tastes (1 if $n_r=1/4$)
for each valence flavor.\footnote{%
Alternatively, as discussed in app.~\protect\ref{app:valence}, one
can average over tastes and multiply by an appropriate overall factor.}
This projection is simple in practice because the staggered propagator
(discussed in app.~\protect\ref{app:PT}) is taste invariant.
It is in this sense (i.e. after projection) that the rooted staggered
PT is physical. 
Note that the projection is only possible for an 
{\em integer} number of tastes.

The upshot is that it is very plausible that, order by order,
PT for rooted staggered fermions (with the ``$n_r=N_f/4$-th'' root) 
reproduces renormalized correlation functions
in the continuum theory with $N_f$ flavors, as long as one
uses either taste projectors or appropriate
averages over tastes in valence loops.
Thus not only the $\beta-$function (which can be accessed using purely
gluonic correlators) but also the anomalous dimensions of the quark 
mass and external operators will be as in the desired continuum theory
(up to the usual dependence on scheme). 
In other words, the unphysical features of rooting
do not appear (or can be removed) in PT, and we find the
same fixed point as with uncontroversial fermions. 
It is important to stress that this holds only when one sends $a\to 0$.
Scaling violations can, and indeed do, contain unphysical effects.

One might be concerned that the argument just given holds for {\em any} $n_r$,
e.g. $n_r=\pi/4$ leading to $N_f=\pi$. It is true that such a theory is
renormalizable, so that its renormalized Euclidean correlation functions should have
a well-defined continuum limit. What is not implied for general $n_r$ is
that the resulting PT is unitary, i.e. that it corresponds to a physical
Minkowski theory. As already noted above,
unitarity will hold only for $n_r=N_f/4$ with $N_f$ a non-negative integer.
Otherwise the resulting PT describes a theory which is irreducibly partially
quenched. Another way of saying this is that only for $N_f$ a non-negative integer
does the corresponding continuum theory have an underlying local Lagrangian,
from which unitarity can be shown.
Note that no claim of unitarity for
the rooted theory for $a>0$ is being made here---the logic is that
one sends $a\to0$ in Euclidean space and finds that the renormalized
correlators are those of a familiar perturbative scheme, which, from
standard continuum considerations, is known to be physical.

Let me pause and emphasize an important conclusion:
\vspace{-.2cm}
\begin{center}
\fbox{\parbox{5truein}{
{\bf It is important to complete the proof of renormalizability for unrooted
staggered fermions, and to confirm the assumptions needed to
extend this result to $\mathbf{n_r=1/4}$.}
}}
\end{center}
In the last year, Giedt has taken a first step towards a proof by
generalizing a power-counting theorem of Reisz to staggered fermions~\cite{Giedt}.

\bigskip
If we accept that rooting gives the correct fixed point in PT are we done?
Does this not imply that the continuum limit of the rooted theory is correct?
{ No!} It is a very important step,
but there are four issues that remain to be resolved.
The first two are not specific to rooted theories,
and are unlikely to be problematic.
I include them for completeness.
\begin{enumerate}
\item
One must be in the domain of attraction
of the desired fixed point. This must be addressed by simulations.
One approach is to study
the scaling of $\alpha_s$ and compare it to the perturbative
predictions, as has been done successfully for pure gauge
theories and for QCD with $N_f=2$ Wilson fermions.
\item
Non-perturbative effects at short distances could
invalidate the perturbative analysis. 
Again, numerical tests of scaling
can rule out (or in) this possibility.
\end{enumerate}
The remaining two issues are specific to rooted theories.
\begin{enumerate}
\item[3.]
The BGS argument implies that the rooted theory is unphysical for any $a>0$,
while the perturbative fixed point is physical. 
There is no technical inconsistency
since the BGS argument relies on {\em non-perturbative} long-distance contributions
to correlation functions.
There is a need, however, to understand how these two results
can be made consistent. There are two possibilities,
one ``good/ugly'', one ``bad'':
\begin{enumerate}
\item[]{\bf Good/ugly:}
The non-locality vanishes as $a\to 0$. As already noted in sec.~\ref{sec:BGS},
this possibility is consistent with the BGS analysis.
\item[] {\bf Bad:}
The PT analysis is not pertinent:
the rooted theory does not run to the perturbative fixed point, and remains
non-local even at $a=0$. This might happen if
the classification of operators into relevant and irrelevant
differs from that in a standard RG framework,
which is possible since there is no 
non-perturbative local single-taste construction of the rooted theory.
Crudely speaking this amounts to saying that
we do not have experience with, or intuition about, theories with
non-local actions, so we should be wary of using standard tools. 
Note that the example from statistical mechanics suggests that standard
tools (here PT) can be used,
but, of course, this lesson may not apply to the theory at hand.

\end{enumerate}
To study this issue we need an extended RG framework which can deal with 
rooting. If the ``good/ugly'' possibility is favored, then
we also need to make plausible a mechanism for the vanishing of
non-locality in the continuum limit.
\item[4.]
We must also consider the rooting problem in the valence sector
in a non-perturbative context. That is, even if the sea-sector
is properly dealt with by rooting, how do we know that the
valence correlation functions that are calculated in practice
give physical results in the continuum limit?
\end{enumerate}
I discuss all of these issues (and more) in the following.

\subsection{Numerical results on scaling}
\label{sec:scaling}

\begin{figure}
\centerline{\epsfxsize=6in\epsfbox{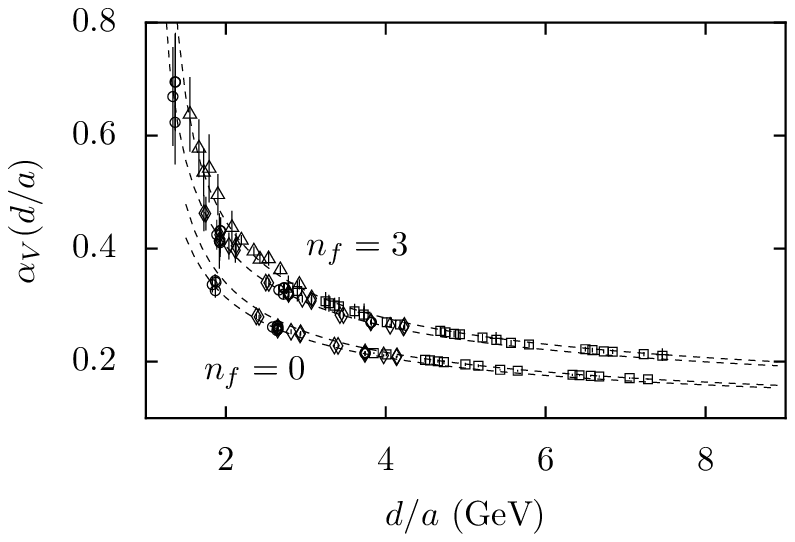}}
\caption{Results for the strong coupling constant,
in the potential scheme, versus scale~\protect\cite{Masonalpha}.}
\label{fig:alpha}
\end{figure}

I begin by briefly showing what is known about the non-perturbative
$\beta-$function with rooted staggered quarks. This has been studied
in Ref.~\cite{Masonalpha} by calculating Wilson loops numerically
and comparing to two-loop (rooted) perturbation theory. This allows
an extraction of the coupling constant at a scale which depends
on the size of the loop. Note that fermion loops do
enter at two-loop order so the perturbative result is sensitive to rooting.
The results are shown in Fig.~\ref{fig:alpha} (with ``$n_f=3$'' referring
to the $2+1$ flavor MILC lattices, and ``$n_f=0$'' to quenched results).
The dashed lines show the predicted
four-loop scaling for the two theories, with initial values chosen
to bracket the simulation results. 

What I want to stress here is that the results in
the perturbative regime (say $d/a \gtapprox 2\;$GeV) are consistent
with the predicted scaling. While this is nowhere near as stringent
a test as has been made for $N_f=2$ Wilson fermions~\cite{Wilsonalpha}, 
using the Schr\"odinger functional definition of $\alpha$~\cite{SchrF}, 
it is nevertheless a non-trivial test that 
QCD with rooted staggered fermions is
governed by the desired fixed point.

\subsection{Renormalization group analysis}
\label{sec:RG}

In this section I describe recent work of Shamir, who
studies rooting using an RG framework~\cite{ShamirRG06}.
The main conclusion is that, given certain assumptions,
the non-locality can be bounded and shown to
vanish when $a\to0$, leaving a theory which is in the
correct universality class. (To make this more precise I
will need to set up the RG framework.)
This is a strong, positive claim for rooted staggered fermions,
and, not surprisingly, a key question is 
the plausibility of the assumptions.
In the following I summarize the argument and critically
examine the assumptions.

\begin{figure}
\centerline{\epsfxsize=6in\epsfbox{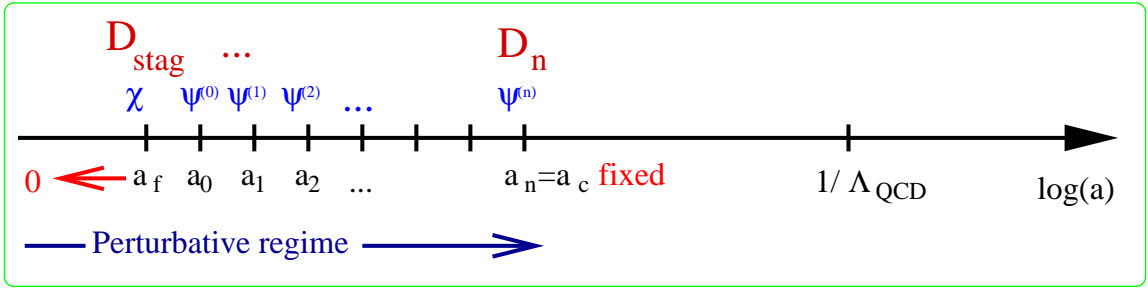}}
\caption{RG framework and notation. Details are discussed in text.}
\label{fig:rg}
\end{figure}

The RG framework is sketched in fig.~\ref{fig:rg}.
One starts on the ``fine'' lattice, with spacing $a_f$, and
blocks $n+1$ times by a factor of 2 until one reaches
the ``coarse'' lattice, with spacing $a_c = 2^{n+1} a_f$.
The aim is to determine the properties
 of the action on the coarse lattice,
whose spacing is held fixed, 
while $a_f$ is sent to $0$ by sending $n\to \infty$.
Applying this procedure to, say, Wilson fermions, 
one would end up with
a so-called perfect lattice action at the coarse scale. The issue
is, of course,  what happens with rooted staggered fermions.
Although it does not matter at this stage, it will be assumed later
that $a_c \Lambda_{\rm QCD}\ll 1$, so that all the blocking steps
are in the perturbative regime.

The fermion blocking is carried out using the ``unrooted'' fields---we
will see later how the results carry over to the rooted theory.
One begins on the fine lattice with the one-component 
staggered field $\chi$, which, in the first step, is transformed
to the Dirac-taste basis (labeled $Q$ earlier, but $\psi^{(0)}$
here, following Shamir). This is not a 
standard blocking transformation,
since the number of fermion degrees of freedom is unchanged. 
The gauge fields are, however, blocked in this first step.
Subsequent blocking steps involve standard thinning of both fermion and
gauge degrees of freedom, with the fermions remaining in the
Dirac-taste basis and labeled $\psi^{(k)}$. 
Although I must skip over most of the details
(such as how gauge and hypercubic invariance are maintained)
I do want to mention that 
fermion blocking is implemented using Gaussian kernels. 
For example, to move from the
$k\!-\!1$'th to the $k$'th fermionic level, one uses
\begin{equation}
\int D\bar\psi^{(k\!-\!1)}D\psi^{(k\!-\!1)}
\exp\{-\alpha_k[\bar\psi^{(k)\dagger} - 
\bar\psi^{(k\!-\!1)\dagger}Q^{(k)\dagger}]
            [\psi^{(k)} - Q^{(k)} \psi^{(k\!-\!1)}]\}
\,,
\end{equation}
with $Q^{(k)}$ a local map implementing gauge covariant 
averaging over a hypercube,
and $\alpha_k$ a parameter conveniently chosen to be $\sim 1/a_k$.

In a standard RG analysis one would now integrate out all levels of fermions
and gauge fields except those on the coarse lattice. This would, however,
lead to multifermion interactions (since the $Q^{(k)}$ include gauge fields),
and one would lose analytic tractability. Instead, Shamir 
keeps all levels of the gauge field while integrating
out all except the coarse-lattice fermions.
All fermionic integrals are then Gaussian and can be done explicitly. 
At each stage one gets a blocked Dirac operator, $D_k$, as well as
an additional contribution to the blocked gauge action.
For present purposes, the key result is
\begin{equation}
\det[D_\stag] \propto \prod_{k=0}^n 
\det[G_k^{-1}]  \det[D_n]
\equiv \prod_{k=0}^n 
\exp(-4 S_{\rm eff}^{(k)}) 
\det[D_n]
\,,
\label{eq:blockeddet}
\end{equation}
where 
$G_k^{-1}$ and $D_n$ are known explicitly:
\begin{equation}
G_k^{-1} = D_{k-1} + \alpha_k Q^{(k)\dagger} Q^{(k)}\,,
\qquad
D_k=\alpha_k-\alpha_k^2 Q^{(k)} G_k Q^{(k)\dagger}\,,\qquad
D_{-1}=D_\stag
\,.
\label{eq:Gkres}
\end{equation}
The proportionality constant in (\ref{eq:blockeddet})
is independent of the configuration.

Thus far this is kinematics:
eq.~(\ref{eq:blockeddet}) is valid configuration by
configuration on any ensemble.
The important dynamical issue is the locality of
the blocked action. 
Integrating out $\psi^{(k-1)}$ must be shown to lead
to an effective gauge action, $S_{\rm eff}^{(k)}=\tr \ln G_k/4$, 
and a blocked Dirac operator $D_k$,
local at scale $a_k$.
This requires both $G_k^{-1}$ {\em and} $G_k$ to be local at this scale
(both appear when taking functional derivatives of
$S_{\rm eff}^{(k)}$ with respect to the gauge field,
and $G_k$ enters in the expression for $D_k$).
In an inductive argument, $G_k^{-1}$ is local because $D_{k-1}$
and $Q^{(k)\dagger}Q^{(k)}$ are, so the real issue
is the locality of $G_k$.
What matters for this are the properties of the 
eigenvectors of $G_k^{-1}$ with small eigenvalues---if there are
no small eigenvalues or if their eigenvectors are localized,
then $(G_k^{-1})^{-1}=G_k$ will be local.
In the free theory there are no small eigenvectors:
where the Dirac operator $D_{k-1}$ is small---i.e. for small
momenta---the $Q^{(k)\dagger}Q^{(k)}$ term is like a large
mass of size $\alpha_k \sim 1/a_k$.\footnote{%
For large momenta, and for $k>0$,
$Q^{(k)\dagger}Q^{(k)}$ is small 
(it is a projection operator) but this
is compensated by the fact that $D_{k-1}$ is large.}
This can break down, however, on rough gauge fields,
because $D_{k-1}$ need not be small.
The situation is similar to that
of Wilson fermions with a large negative mass,
for which we know that the derivative can become large
on rough gauge configurations.
Clearly a more sophisticated discussion is needed.

That given by Shamir is drawn from studies of
the ``mobility edge'' of the hermitian Wilson-Dirac operator
with a large negative mass~\cite{mobilityedge}.
This operator does have small eigenvalues (indicating 
that the derivative term is large) but in a suitable region
of the phase diagram these are
{\em exponentially localized at the scale of the lattice spacing}, 
so that the contribution 
of the corresponding eigenvectors to the inverse remains local.
Only eigenvalues above the mobility edge have delocalized eigenvectors,
and as long as this edge remains $\gtapprox 1/a$, their contribution
to the inverse will also be local.
Shamir argues that this picture holds also for $G_k^{-1}$
(or, more precisely, for the
hermitian form $H_k=(\gamma_5\otimes\xi_5) G_k^{-1}$, 
which is sufficient to establish locality),
irrespective of whether the ensemble is rooted or unrooted,
as long as one is in the weak coupling regime.
Weak coupling enters because numerical results indicate
that the mobility edge lies close to the perturbative value,
$\sim 1/a$, in this regime.
The irrelevance of rooting is based on the fact that the
gauge action dominates the measure for the rough gauge fields.

The outcome is the {\bf first key assumption}: 
that $S_{\rm eff}^{(k)}$ and $D_k$ are local at scale $a_k$,
irrespective of whether the ensemble is rooted or unrooted.
I find the argument for this assumption, sketched above,
to be plausible. It is ``anchored'' by the unrooted case:
here the theory is local on the fine lattice, and blocking
by a local RG transformation should keep it local at each stage,
for this is a standard RG application.
This implies that Shamir's mobility edge argument must work for
the unrooted theory.

Accepting the assumption, what have we learned about the
rooted theory? The fermion weight factor is then
\begin{equation}
%
%
\det[D_\stag]^{1/4}
 \propto \prod_{k=0}^n 
\det[G_k^{-1}]^{1/4}
\det[D_n]^{1/4}
= \prod_{k=0}^n 
\exp(-S_{\rm eff}^{(k)}) 
\det[D_n]^{1/4}
\,.
\label{eq:Seffdeff}
\end{equation}
The assumption implies that the effective gauge action introduced
by blocking, here $S_{\rm eff}^{(k)}$ 
(without the factor of 4), remains local on the rooted ensemble.
Thus the non-locality in the one-taste theory, which we know
is present from BGS, must result from
the $\det[D_n]^{1/4}$ term. This means that, crudely speaking,
the non-locality resides in the IR.
This is consistent with my earlier discussion of PT---no
unphysical behavior should appear in the perturbative regime---but
goes beyond it, by indicating how non-localities from
possible non-perturbative effects at short distances
(rough gauge fields) can be controlled.

Now I move to the second part of the argument: 
bounding the effects of non-locality in $\det[D_n]^{1/4}$.
This is intertwined with the issue of taste breaking, since
it is such breaking that precludes taking
the fourth-root of the fermion determinant.
It thus makes sense to decompose the blocked Dirac operator as
\begin{equation}
D_n = 
\underbrace{
\widetilde D_{n}\otimes \mathbf{1}}_{\textrm{taste invariant}} 
+ \underbrace{\Delta_n}_{\textrm{taste-breaking}} 
\,.
\label{eq:Dndecomp}
\end{equation}
Recall that, given Shamir's first assumption, $D_n$ is local
on the rooted ensemble, and thus so are the two parts on
the r.h.s., as they are obtained by taste-projection.
The aim now will be to argue for the
{\bf second key assumption}: $|\Delta_n|$ is bounded from above
and vanishes as $a_f\to 0$ (i.e. as $n\to\infty$).
This is to be contrasted to $D_n$ itself, which is bounded from below
by $m(a_c)$, the quark mass at scale $a_c$, so that, as long as we
work at non-zero $m$, the taste invariant part dominates as $a_f\to 0$.
In particular, for small enough $a_f$, the rooted determinant 
\begin{eqnarray}
\det[D_n]^{1/4} &=& 
\det[\widetilde D_{n}\otimes {\bf 1}]^{1/4}
\ \det[1 + \Delta_n(\widetilde D_{n}\otimes {\bf 1})^{-1}]^{1/4} \\
&=&
\det[\widetilde D_{n}]
\ \exp\{
(1/4) \tr\ln
[1 + \Delta_n(\widetilde D_{n}\otimes {\bf 1})^{-1}]
\}
\label{eq:rooteddetcont}
\end{eqnarray}
can be legitimately expanded in powers of $\Delta_n$.
The first factor in (\ref{eq:rooteddetcont}) is the determinant
for a local one-taste theory with Dirac operator
$\widetilde D_{n}$, while the second
shows explicitly [through the inverses
of $(\widetilde D_{n}\otimes {\bf 1})$]
the non-locality that we knew had to be present.\footnote{%
Similarly, one can see, by taking
the fourth power of (\protect\ref{eq:rooteddetcont}),
how a non-local ``gluonic'' effective action can lead to
taste breaking in the fermion sector
of the unrooted staggered theory, as advertised 
in sec.~\protect\ref{sec:BGS}.}
The key point, however, is that if $|\Delta_n|$ vanishes
as $a_f\to 0$ then the non-locality also vanishes in that limit
(the second factor becoming unity).

An important property of $D_n$ that has been used in this
construction is that its taste singlet part, $(\widetilde D_n\otimes {\bf 1})$,
has no doublers~\cite{BGS,ShamirRG06}.
For this to be true,
at least one blocking step must be taken,
since it does not hold for the original action, $D_\stag$.
This can be seen from eq.~(\ref{eq:SstagQ}):
projecting onto the taste-singlet part reintroduces the doublers.

The local theory we obtain as $a_f\to 0$ contains
a single taste (per flavor) with action
$\sum \bar q \widetilde D_{n\to\infty} q$,
but has the unusual feature of having all the
blocked gauge fields present.
Integrating out all except the coarse lattice gauge fields
will yield a complicated action involving multifermion terms.
This complication is not, however, a theoretical problem.
What matters is that the resulting action is
local at scale $a_c$, and has the same symmetries as,
say, lattice QCD with overlap fermions---flavor and
the symmetries of the blocked lattices 
(hypercubic and translation invariance).
It is thus in the same universality class as QCD.
%

I have been rather cavalier with the limit $n\to\infty$.
This is addressed by Shamir by introducing ``reweighted'' theories
in which one replaces
$\det[D_n]^{1/4}$ with
$\det[\widetilde D_{n}]$ for {\em finite} $n$.
Now one can integrate out a finite number of blocked gauge
fields to obtain a complicated but local single-taste action,
which is manifestly in the QCD universality class.
One can further show, given Shamir's second assumption,
 that the difference between this reweighted theory
and rooted staggered fermions vanishes as $a_f\to 0$.
In other words, there is a sequence of local theories
labeled by $n$, all in the QCD universality class,
which, as $n\to\infty$,
become closer and closer to the rooted staggered theory
having the same fine lattice spacing $a_f$.
Working with these reweighted theories allows
one to avoid the possibility, raised at the end of the previous
subsection, that non-localities can somehow invalidate
the perturbative analysis of RG scaling.

I now return to sketch the argument leading to
the assumption that $|\Delta_n|$ 
is bounded and vanishes as $a_f\to 0$. 
There are two ways in which taste is broken
and correspondingly two parts to the present argument.\footnote{%
The following discussion differs substantially
from that in my talk, due to an improved
understanding developed in
discussions with C. Bernard, M. Golterman
and Y. Shamir.
A more thorough discussion appears in
Refs.~\cite{BGSlat06,ShamirRG06}.
}
The first is that fine lattice gauge fields with momentum components
$\sim \pi/a_f$ lead to taste violating quark-antiquark-gluon
vertices. This is very well known,
and an explicit example is given in app.~\ref{app:PT}. 
What is peculiar about the present RG set-up is that such
gluons are not integrated out, and so can lead
to taste violations in $D_n$, since (tracing back the RG
transformations) this depends on the gauge fields on all
but the coarse lattice. Such contributions to $\Delta_n$
are suppressed, however, by the gluon functional weight.
In perturbation theory the gluons are far off-shell,
and when they are finally integrated out, lead to taste-violating
four-fermion interactions, explicitly suppressed by 
$1/k^2 \propto a_f^2$ up to 
logarithms~\cite{Susskind,Lepage,Lagae,LeeSharpe}.

The second source of taste violation is that present 
already in the second term in the
free theory action (\ref{eq:SstagQ}).
The lowest dimension operator in this term is
$\sum_\mu \bar Q (\gamma_5\otimes \xi_\mu\xi_5) \partial_\mu^2 Q$,
which has dimension 5 and thus comes with a factor of $a_f$.\footnote{%
It is probable that taste breaking from this source
is actually absent at $O(a_f)$. 
In the unrooted theory one can argue
using the Symanzik improvement program that there are no
$O(a)$ corrections with staggered fermions~\cite{SSimpstag,Luo1}. 
This implies that there is a field redefinition which removes
the $O(a)$ term in eq.~(\protect\ref{eq:SstagQ}), even in the
interacting theory, as discussed by Luo~\cite{Luo1}.
Since this is an essentially kinematic result, it is plausible
that it carries over to the rooted ensemble.
Taste violation is then moved to higher order in $a_f$.
Since I am not making use of the dependence on $a_f$,
I do not pursue this point further.}
Shamir argues that, as the fermions are integrated out,
the scaling of the contributions of the covariant form of
this operator will have the standard form---the overall factor of $a_f$
will remain, up to logarithmic corrections, and
$\partial_\mu^2$ gives $\sim 1/a_c^2$.
Standard scaling is expected because, from Shamir's first assumption,
the interactions are local in the range from $a_f$ to $a_c$.

The arguments leading to the assumption that $|\Delta_n|\to 0$
are the least well developed part of Shamir's approach.
It is important to check that there are no surprises
when one develops perturbation theory for the non-standard RG
setup. 
Nevertheless, even without this check,
I think the conclusion is plausible.
This is again 
because of the ``anchor'' provided by the unrooted theory,
together with the plausible claim that ``rooting works'' in PT
(as discussed in sec.~\ref{sec:PT}),
and the fact that the RG running is in the perturbative domain.
For a more thorough discussion see Ref.~\cite{BGSlat06}.

In summary, Shamir's analysis provides a RG framework
which is able to deal with the rooted determinant
despite the lack of a local one-taste theory.
Accepting two key assumptions,
one can see how the non-locality implied by BGS
can be present for non-zero lattice spacing and yet
vanish in the continuum limit.
This is the ``good/ugly'' option from the previous sub-section.
Although the RG framework is non-standard, 
it is plausible that the analysis
of operators in PT into relevant, marginal and irrelevant
still holds, thus making the ``bad'' option unlikely.

Clearly more work would be welcome, in particular
to test the key assumptions. Both can be tested
numerically, and the second assumption can also be
tested in PT.

\subsection{Understanding non-locality using chiral perturbation theory}
\label{sec:ChPT}

Effective chiral theories turn out to provide an alternative
and powerful avenue to tame the non-locality.
In this section I describe the recent
work of Bernard~\cite{CB06}, who has shown the following result.
\vspace{-.2cm}
\begin{center}
\fbox{\parbox{5truein}{
\textbf{
If partially quenched staggered chiral perturbation theory 
(\schpt) is a valid effective field theory (EFT) for
unrooted but partially quenched staggered fermions,
and accepting some plausible technical assumptions,
then the correct EFT
for rooted staggered fermions is ``\schpt\ with the
replica trick'' (r\schpt).
}}}
\end{center}
Here the EFTs describe the physics of PGBs---their interactions
and those with other, heavier, hadrons.
Bernard's result is clearly important, and it was
a surprise to me that one could derive such a strong statement.
I will attempt here to get to the essence
of the derivation without going into too much detail.

The implication of the result is that we know the
correct EFT describing the long-distance physics of rooted 
staggered fermions at small, but non-zero, lattice spacings, 
namely r\schpt~\cite{LeeSharpe,rSChPT}. 
I stress that the lattice spacing is an
explicit parameter in this EFT, with discretization errors
introduced following the method of Ref.~\cite{ShSi},
using Symanzik's effective Lagrangian~\cite{Symanzik}.
R\schpt\ has two key properties.
First, it is unphysical for $a\ne 0$, exhibiting, 
for example, violations of unitarity.
Second, the results go over to those of standard \chpt\ for QCD
when $a=0$, as long as one considers the appropriate external states.
(The latter point is discussed further in sec.~\ref{sec:diseases}.)
These are exactly the properties that must hold in order
both to be consistent with the BGS argument and obtain the correct
continuum limit.
What is new here is not that r\schpt\ has these properties,
but rather that this EFT---with the apparently {\em ad hoc} feature
of adding factors of $1/4$ by hand to represent rooting in the underlying
theory~\cite{rSChPT}---can be shown to be correct.
 
The importance of this result is twofold. First,
it establishes that the continuum limit of the PGB physics in the
rooted staggered theory is that of QCD
(modulo one caveat about low-energy coefficients [LECs] discussed below).
Second, it ``tames'' the non-locality which is present away from the continuum
limit: the unphysical consequences are known explicitly
for the  properties of PGBs and their interactions with each
other and with other hadrons. Perhaps ``known'' is too strong a term:
the unphysical features can be parameterized in terms of LECs
in the EFT. In any case, the fitting that has been done
using r\schpt, which has proven essential in practice, is now
on a much firmer theoretical footing.

I have emphasized in my formulation of Bernard's result the
need to assume the validity of PQ\chpt.
I have done so both because I think this is the least well
established assumption in his analysis,\footnote{%
Here I differ from Bernard's presentation, which
emphasizes the need for two technical assumptions.
These are mentioned below,
but I do not have time to discuss them in detail. 
I find both assumptions very plausible.}
and because it helps one understand the result.
Let me discuss these two points in turn.
What is the status of PQ\chpt?
PQ theories are unphysical, and the arguments used to
justify EFTs in general~\cite{Weinberg} and \chpt\ in particular~\cite{Leutwyler}
only partly apply. One has to make additional assumptions, as discussed
in Refs.~\cite{ShSh2,Shlec06}. Nevertheless, 
I do not see any reason at present to doubt
the validity of PQ\chpt: it appears to correctly represent the
unphysical nature of the underlying PQ theory, and is supported
by a growing body of numerical evidence 
(an example of which was noted in sec.~\ref{sec:update} above).
I am partly stressing this point to encourage further study
of the issue.

Now I return to the unexpected nature of the result:
how is one able to determine the EFT
for an underlying theory which is unphysical?
The key points are that the rooted theory
has a precise formulation as a functional integral, and that,
crudely speaking,
one is able to trade rooting for partial quenching.
While this is superficially just trading one unphysical feature
for another, it is actually significant progress.
The validity of PQ theories is a general issue,
one that can be studied {\em with any fermion discretization}
and also in the continuum limit,
whereas rooting is, in practice, staggered-specific.
Furthermore,
the underlying PQ theory has a local Lagrangian formulation
(albeit involving ghosts) which is not available for the rooted theory.

To explain these comments in more detail I now present my distillation
of Bernard's argument.
Consider the partition function for four flavors 
of rooted staggered fermions with non-degenerate masses:
\begin{eqnarray}
Z^{\rm root}(m_1,m_2,m_3,m_4) &=& 
\int d\mu_g \left\{\prod_{i=1}^4 \det[D_\stag(m_i)]
\right\}^{1/4}\\
&=&
\int d\mu_g \left\{\prod_{i=1}^4 \exp[{\frac14}\tr\ln D_\stag(m_i)]
\right\}
\,,
\label{eq:rooted4flavor}
\end{eqnarray}
where $d\mu_g$ is the gluonic functional measure (in which resides,
implicitly, the strong coupling constant).
There are two particularly interesting choices of quark masses.
First, setting all the masses equal, one obtains the theory
with one {\em unrooted} staggered flavor:
\begin{equation}
Z^{\rm root}(m,m,m,m) =
\int d\mu_g \det[D_\stag(m)] \equiv
Z^{\rm stag}(m)
\,.
\end{equation}
This is a physical theory which anchors the analysis.
Second, one can obtain the ``MILC'' partition function as follows:
\begin{equation}
Z^{\rm MILC}(m_\ell,m_s) =
Z^{\rm root}(m_\ell,m_\ell,m_s,\sim 1/a)
\,.
\end{equation}
Sending one of the quark masses to the lattice cut-off 
does lead to a renormalization of the coupling constant 
(and thus to a shift in the lattice spacing), but this
is not important for the issue at hand, namely the
determination of the form of the EFT.

Clearly one can move from $Z^\stag$ to $Z^{\rm MILC}$ by varying
the quark masses. The strategy of the argument is to do so
by calculating all derivatives of $Z^\stag$ w.r.t. the $m_i$
at the degenerate point, and then summing the series.
As shown by Bernard, this can be done within the EFTs.
Using this method, one is able to start from a physical
theory ($Z^\stag$) with known EFT and obtain the EFT for $Z^{\rm MILC}$.
The simple words ``summing the series'' summarize the main
technical analysis in Ref.~\cite{CB06}, which I will not reproduce.
The two technical assumptions needed are
(i) that summing the series gives
the complete answer---i.e. that there are no non-analyticities, which
is plausible since all $m_i$ are greater than zero at all stages
(so that all particles are massive);
and (ii) that there are no unexpected issues in the EFT when one decouples
the fourth quark ($m_4\to \sim 1/a$).

The derivatives one needs are illustrated by the following simple example:

\begin{minipage}{0.4\textwidth}
$\ \ \ \ C(x,y)=
\frac{\partial^2 \ln Z^{\rm root}}{\partial m_4(x) \partial m_4(y)}\bigg|_{m_i\ \textrm{equal}}=$
\end{minipage}
\begin{minipage}{0.5\textwidth}
\begin{center}
\includegraphics[width=0.9\textwidth]{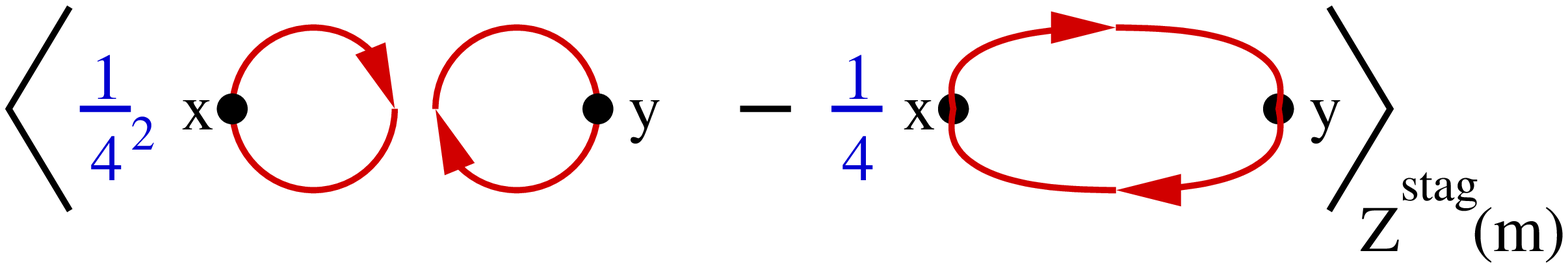}
\end{center}
\end{minipage}

\noindent
Here the lines indicate propagators---elements of $D_\stag^{-1}$.
To obtain such results one must treat the quark masses as space-time
dependent sources. Inverses of the {\em unrooted} $D_\stag$ appear
because of the form of eq.~(\ref{eq:rooted4flavor}),
with rooting leading to the factors of $(1/4)^{\#\ \rm loops}$. 
Although the expression on the r.h.s. is evaluated in the unrooted
ensemble, $Z^\stag(m)$, it is not a physical combination of correlators,
as it cannot be obtained by taking derivatives w.r.t. the common mass $m$.
Such derivatives lead to the wrong relative factors of $1/4$ between the two terms.
The expression on the r.h.s. is in fact a quantity in a PQ theory with partition
function
\begin{eqnarray}
\lefteqn{Z^{\rm PQstag}(m,M_{V},\widetilde M_{V}) = 
\int d\mu_g \det[D_\stag(m)] \times} 
\nonumber\\
&&
\prod_{i=1}^{N_V}
{ D\bar\chi_{V_i} D\chi_{V_i}}
{D\widetilde\chi_{V_i}^\dagger D\widetilde\chi_{V_i}}\ 
\exp[-\underbrace{ \bar\chi_{V_i} D_\stag(M_{V,ij}) \chi_{V_j}}_{
\textrm{valence quarks}} \!- \!
\underbrace{
\widetilde\chi_{V_i}^\dagger D_\stag(\widetilde M_{V,ij}) \widetilde \chi_{V_j}}_{
\textrm{ghost quarks}}
]\,,
\end{eqnarray}
with the explicit rewriting of $C(x,y)$ being
\begin{equation}
\frac{\partial^2 \ln Z^{\rm root}}{\partial m_4(x) \partial m_4(y)}\bigg|_{m_i=m}
=
{\frac1{4^2}}
\frac{\partial^2 \ln Z^{\rm PQstag}}{
\partial M_{V,11}(x) \partial M_{V,22}(y)}\bigg|_{M_V=\widetilde M_V=m}
\!\!+
{ \frac1{4}}
\frac{\partial^2 \ln Z^{\rm PQstag}}{
\partial M_{V,12}(x) \partial M_{V,21}(y)}\bigg|_{M_V=\widetilde M_V=m}
\,.
\end{equation}
This illustrates how rooting can be traded for partial quenching.
The need to add factors of $1/4$ by hand in the PQ theory betokens
the appearance of such factors in r\schpt.

Similar identities hold for any number of derivatives.
In Bernard's analysis they are applied to the EFTs, i.e. to the
partition functions which generate the long-distance parts of correlation
functions. By assumption, we know the EFT for $Z^{\rm PQstag}$---it is 
the partially quenched version of the ``original'' staggered
\chpt\ of Lee and myself~\cite{LeeSharpe}. 
By evaluating the derivatives in the PQ theory, and summing the
multivariate Taylor series, one can construct the EFT describing
the rooted staggered theory.
As announced, the result is r\schpt~\cite{rSChPT}.\footnote{%
In the literature, this is referred to as simply \schpt,
without the ``r'', because in practice one only uses
staggered chiral perturbation theory in the presence of rooting.
Here I keep the ``r'' for the sake of clarity.}

Now I return to the caveat mentioned above.
Bernard's argument as given establishes that the
EFT has the same {\em form} as r\schpt, but not that the
LECs in it have the same values as those in 3 flavor \chpt.
In other words, it remains possible that the \chpt\ obtained when
$a\to 0$, while having the same form as that for continuum QCD,
has different LECs. To show that the LECs are those of QCD \chpt\
one must, at present, rely on the RG argument of the previous
section to demonstrate that the continuum theory is QCD 
and not just QCD-like.\footnote{%
An alternative approach was discussed here by BGS in which
the EFT is obtained more directly using the reweighted local
theories mentioned in the previous section~\cite{BGSlat06}.}
It would be clearly better to avoid dependence on the RG argument,
and I think this should be possible. The theoretical issue here
has nothing to do with rooting. For example, it does not arise in the
4-flavor rooted theory, where Bernard's arguments
show that the LECs in the corresponding
r\schpt\ are indeed those of the EFT for 
the QCD-like theory with 4 light flavors. The issue is what happens
to the LECs when one decouples the fourth flavor.

In summary, the EFT approach gives both another line of
argument showing that the continuum limit of rooted staggered
fermions is correct (in the PGB sector) and provides
explicit formulae with which to do the chiral and continuum
extrapolations. It is complementary to the RG approach,
which is general (not restricted to the PGB sector
or the chiral regime) but does not provide explicit formulae.

\section{Diseases of rooted staggered fermions?}
\label{sec:diseases}

The title of this section is adapted from that of a contribution
to this conference by Creutz~\cite{Creutzlat06}.
This is based on a recent paper~\cite{Creutz06}
in which he argues that rooted staggered fermions have
fundamental problems related to having too many symmetries.
In the terminology of my introduction, Ref.~\cite{Creutz06}
argues that rooted staggered fermions are ``bad''.
It is by now clear, as shown in a rebuttal by
Bernard, Golterman, Shamir and myself (BGSS)~\cite{BGSS}, that
the arguments of Ref.~\cite{Creutz06} do not demonstrate this.
What they do highlight, however, is the delicate nature of
the continuum limit of the rooted staggered theory, 
and the need to work at non-zero quark masses while taking
this limit.
The issue is then whether one should view this situation 
(which I will explain in the following) as
a ``disease'' or rather an ugly, but tolerable, feature.

The essence of Creutz's concern is that the $D_\stag$ residing
in the rooted determinant retains the symmetries of
the unrooted theory. In particular, it retains the
softly broken $U(1)_\epsilon$ symmetry.
In the single-component basis this rotates the quark mass $m$
into $m \exp[i\epsilon(n) \theta_\epsilon]$, where
$\epsilon(n)=\pm 1$ for even/odd sites.
Thus when the rotation angle is $\theta_\epsilon=\pi$ 
the quark mass changes sign.
It follows that $\det[D_\stag(m)]=\det[D_\stag(-m)]$
on all configurations,
and thus that the rooted determinant 
is independent of the sign of $m$,
\begin{equation}
{\det}^{N_f/4}[D_\stag(m)] = {\det}^{N_f/4}[D_\stag(-m)]
\,.
\label{eq:rootediseven}
\end{equation}
Here I have generalized to a theory which, if rooting is valid,
gives $N_f$ flavors in the continuum limit, and made use of
the fact that the positive root is taken by definition.
Equation (\ref{eq:rootediseven})
implies that the partition function of the rooted theory
is an even function of $m$ {\em for any $N_f$}.
Now consider the theory in a finite (but arbitrarily large) volume
and for $a\ne 0$, i.e. before any limits have been taken.
Then the fermionic weight is a finite polynomial in $m$ and 
(since all eigenvalues have non-zero imaginary parts,
except possibly on a measure-zero set of configurations) can have
no non-analyticities for real $m$. This implies that the partition function
depends on $m^2$ (rather than the even, but non-analytic, function $|m|$).
Again this is true for all $N_f$.

Now we come to the apparent contradiction noted by Creutz. If $N_f$ is
odd, e.g. $N_f=1$, then we know that the continuum theory has no $m\to -m$
symmetry, because the $U(1)_A$ transformation that could change the sign of $m$
is anomalous. Thus the continuum partition function
is expected to be a general function of $m$, 
containing both even and odd powers. To make this less formal we can
consider the lattice theory with overlap fermions, for which the
same considerations hold for any lattice spacing.
Thus we are led to the problem: how can the rooted staggered theory,
whose partition function is even in $m$ for any non-zero $a$,
have a continuum limit without this symmetry?

The escape from this apparent paradox is fairly simple, and depends on
two facts. First, that, by construction, rooting gives a continuum theory
with a positive mass, $|m|$, irrespective of the sign of the bare lattice mass.
And, second, that non-analyticities can occur since we are taking a limit,
$a\to 0$, in which the number of degrees of freedom goes to infinity.
Thus it is perfectly possible for the limit to produce a general function
of $|m|$. This can then match the continuum form for positive $m$.

This shows that there is no logical inconsistency with rooted
staggered fermions, but we would clearly like more, namely
an argument showing how the non-analytic form $|m|$ can appear.
BGSS argue as follows. 
We know that odd powers of $m$ arise in the continuum
theory from zero modes of $D_{\cont}$, since these are unpaired.
The simplest example is the sector with unit topological charge ($\nu=1$), for which,
on any configuration, and considering only $N_f=1$ for brevity,
\begin{eqnarray}
\det[D_\cont(m)] &=& m \; F_\cont(m^2) \qquad\qquad\qquad (\nu=1)
\\
F_\cont(m^2) &=& \prod_{\lambda>0} (i\lambda +m)(-i\lambda+m)
=\prod_{\lambda>0} (\lambda^2+m^2) > 0
\,,
\end{eqnarray}
where the $\lambda$ are real, and the pairing is due to the $U(1)_A$
transformation.
This formal expression, and those that follow, can be regulated using
overlap fermions, for which one can define topological charge, and
an index theorem, for $a\ne 0$.
The spectrum of $D_\stag$ will differ from that of $D_\cont$ in two essential ways.
First, each eigenvalue of $D_\cont$ will be replaced by a quartet,
split by an amount which is expected to vanish in the continuum limit.\footnote{%
Whether this splitting is expected
to vanish  as $a$ or as $a^2$ is not clear.
This is discussed briefly below,
and in more detail in Ref.~\cite{GSSlat06}.
All I need here is that it vanishes, so I use
the weaker assumption of $O(a)$ splittings.}
The pairing of eigenvalues with opposite signs of the imaginary
part still holds, now because of the $U(1)_\epsilon$ symmetry.
In this way $F_\cont(m^2)$ is replaced by 
$F_\stag(m^2)^{1/4}$---both even, positive functions of $m$.
The second difference between the spectra
is that the zero mode of $D_\cont$
will become a quartet of {\em non-zero} modes,
arranged into two pairs because of $U(1)_\epsilon$.
There are no exact zero modes 
because staggered fermions have no index theorem.
Thus we expect
\begin{equation}
\left\{\det[D_\stag(m)]\right\}^{1/4} = 
\left\{[(\lambda_1^\stag)^2+m^2][(\lambda_2^\stag)^2+m^2] F_\stag(m^2)\right\}^{1/4}
\,,
\end{equation}
with $\lambda_{1,2}^\stag \propto a$.
Note that this is a manifestly even function of $m$, as expected from
the argument above.
If we now take $a\to0$, however, we obtain the claimed non-analyticity:
\begin{equation}
\left\{[(\lambda_1^\stag)^2+m^2][(\lambda_2^\stag)^2+m^2]\right\}^{1/4}
{\lower0.6ex\hbox{$\longrightarrow \atop {a\to 0}$}}\ 
|m|
\,.
\label{eq:gettingnonanalytic}
\end{equation}

This discussion shows how the desired continuum
limit can be achieved if, on the dominant gauge configurations,
the eigenvalues tend to a particular distribution as $a\to 0$
(degenerate quartets, with the expected numbers of near-zero modes).
Numerical studies of eigenvalues find results consistent with 
this behavior~\cite{quartets}.
While this is an important check of rooted simulations, it will
not convince a skeptic, since one does not obtain a physical
understanding of the observed behavior.
For this one has to rely indirectly on the arguments of the
previous section: if the continuum limit is as claimed, then
the eigenvalues must behave as outlined above.

It is clear from eq.~(\ref{eq:gettingnonanalytic}) that,
as one reduces the quark mass, one must work
at correspondingly smaller lattice spacing in order to maintain the
correct mass dependence (up to whatever tolerance one is aiming for).
This point, stressed long ago by Smit and Vink~\cite{SmitVink},
is brought out more clearly by considering the condensate.
If $N_f=1$, the condensate is non-vanishing due to the linear
contribution to the partition function from the $\nu=1$ zero mode
\begin{equation}
\langle\bar\psi\psi\rangle_\cont = - \frac{1}{Z_\cont V} 
\frac{\partial Z_\cont(m)}{\partial m}
{\lower0.6ex\hbox{$\longrightarrow \atop {m\to 0}$}} \ \textrm{non-zero constant}
\,.
\end{equation}
This result holds in any volume, $V$,
unlike the condensate in QCD, which vanishes in the chiral limit
unless one sends $V\to\infty$ first.
This difference arises because the $N_f=1$ condensate is not an order parameter,
since the $U(1)_A$ symmetry is anomalous.
The $N_f=1$ condensate is also a general function of $m$.
By contrast, the condensate with rooted staggered fermions is an {\em odd}
function of $m$, since it is obtained by differentiating 
the even quantity $Z_\stag$.
The contribution from near-zero modes dominates as $m\to 0$ and is
\begin{equation}
\langle\bar\chi\chi\rangle_\stag 
\propto
\frac{m [(\lambda_1^\stag)^2 + (\lambda_2^\stag)^2 + 2m^2]}
{\left\{[(\lambda_1^\stag)^2 + m^2][(\lambda_2^\stag)^2 + m^2]\right\}^{3/4}}
\,.
\end{equation}
In this case [unlike in eq.~(\ref{eq:gettingnonanalytic})] the
limits $a\to 0$ and $m\to 0$ do not commute.
If one sends $a\to 0$ first one obtains $\textrm{sign}(m)$,
which gives a non-zero constant when subsequently sending $m\to 0^+$.
This is the correct order of limits to obtain the physical answer.
By contrast, sending $m\to 0$ first gives zero.

\begin{figure}
\centerline{\epsfxsize=6in\epsfbox{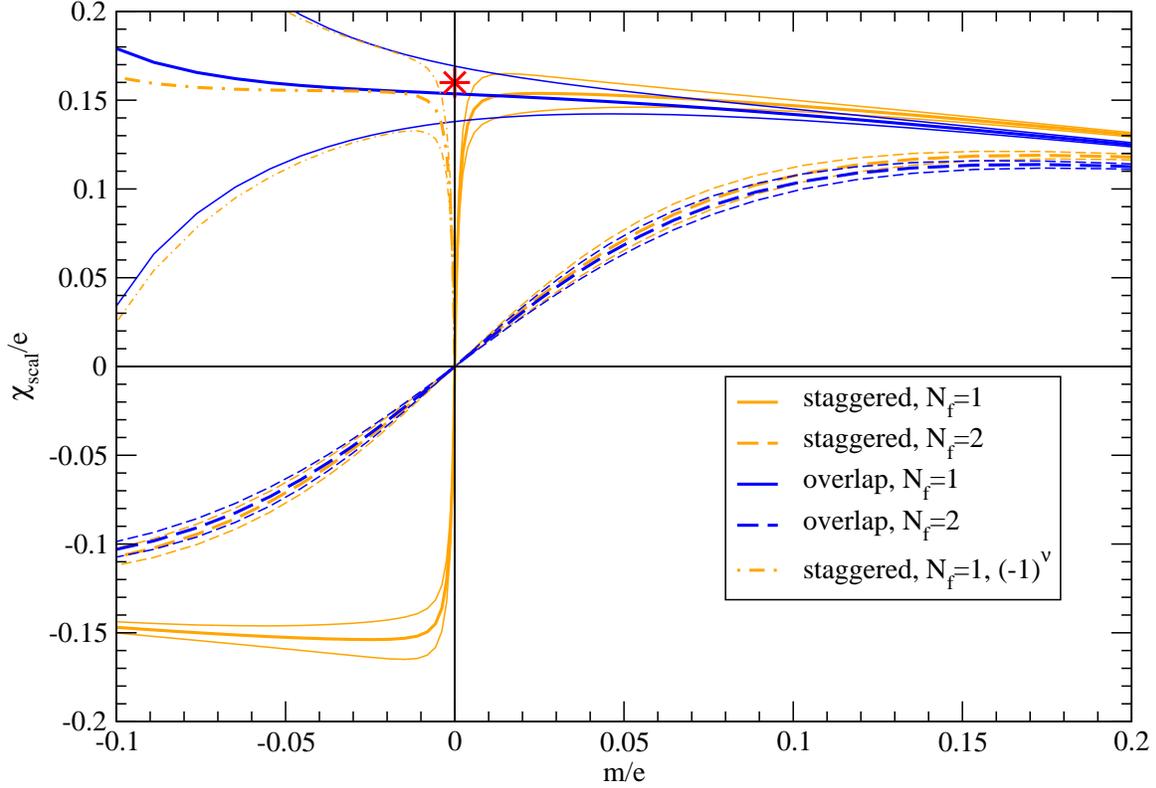}}
\caption{Condensates in the Schwinger model plotted versus quark mass
in dimensionless units~\protect\cite{DH}. 
Results for $N_f=1$ and $2$ are shown, using both
(rooted) staggered and overlap fermions. Each result shows a central
value and a 1-$\sigma$ error band. The asterisk shows the known continuum
result for the $N_f=1$ massless theory. The ``$(-1)^\nu$ curve is
explained in the text.}
\label{fig:DH}
\end{figure}

As above, this is a scenario which shows how rooted staggered fermions can
have the correct continuum form, but does not itself demonstrate that
the resulting condensate is correct. For this one needs other arguments
(those of the previous section) or numerical tests. A very nice test has
been carried out by D\"urr and Hoelbling using the Schwinger model~\cite{DH}. 
In this 2-dimensional model analogous issues arise when considering the
{\em square-}rooted two-taste staggered determinant, for which the
putative continuum limit has a single flavor. The rooted results can be
compared to those obtained using a single-flavor overlap determinant, allowing
a comparison at non-zero lattice spacing. 
This comparison is shown in Fig.~\ref{fig:DH}. The rooted staggered result
tracks the overlap result for positive masses (the difference at larger $m/e$
being understood as a higher order discretization error) until, at $m/e\approx 0.01$,
it plummets to zero (as required by antisymmetry and analyticity).
This form is consistent with the BGSS explanation,
and illustrates the practical implications
of the lack of commutativity of $a\to0$ and $m\to0$ limits.\footnote{%
There is no similar issue for $N_f=2$, where the staggered determinant is unrooted.
The staggered and overlap results show the expected agreement (and oddness in $m$).
This is analogous to the $N_f=4$ case in four dimensions.}

As already noted, rooted staggered fermions always correspond to 
a positive quark mass (irrespective of the sign of the mass in
$D_\stag$). One way to obtain a negative quark mass is to add
a $\theta F \widetilde F$ term by hand, and set $\theta=\pi$.
This is equivalent to adding a weight term $(-1)^\nu$ to the determinant.
While this might be numerically challenging for QCD (due to the sign
problem) it is tractable for the Schwinger model, as shown by the
curved labeled ``staggered, $N_f=1$, $(-1)^\nu$''. 
This nicely matches onto the overlap results for negative mass
(though both have
errors which blow up as $|m|$ increases due to the sign problem).
Combining this result with that for $m>0$ one reproduces the
overlap curve except for a small gap around $m=0$.
The BGSS analysis predicts that this gap will shrink to zero width
as $a\to0$.

How does this analysis impact rooted staggered QCD (with $N_f=3$)?
If one works in the isospin symmetric limit, with
common mass $m_\ell$, and with fixed physical
strange quark mass, then the limits $m_\ell\to0$ and $a\to 0$ commute
for most physical quantities~\cite{CBlimits}. 
The issues described above arise when a single quark mass vanishes,
e.g. when $m_u\to 0$ with $m_d$ and $m_s$ fixed and positive.
The ``staggered, $N_f=1$'' curve in Fig.~\ref{fig:DH} would then
qualitatively describe $\langle\bar u u\rangle$ as a function of $m_u$.
The resulting ``dip'' at $m_u=0$ is certainly an unphysical feature of rooted
staggered fermions. If one wanted to study QCD in this regime one
would have to send $a\to 0$ before $m_u\to 0$, which would clearly 
be numerically challenging. 
Fortunately, this is not a problem in practice for QCD, 
since it is now almost certain that $m_u/m_d$ is not close to zero,
so one does not come close to the ``dip''.\footnote{%
To access the $m_u < 0$ regime (with its interesting phenomenon of
spontaneous CP violation~\cite{CP}) would require an explicit $\theta=\pi$ term
added to the action, analogous to that used in Ref.~\protect\cite{DH}.}

The presence of the dip---which is a definite
conclusion of Creutz's analysis---provides an interesting challenge.
It should be possible to understand its properties using r\schpt\
if this is the correct EFT for the rooted theory. 
The issue is not that $\langle\bar u u\rangle$ is odd
and thus has a dip---this
follows from the same symmetries in r\schpt\ as it does at the
quark level---but rather that the width of the dip vanishes with $a$.
To paraphrase Creutz, how can the UV quantity $a$ enter
into what is apparently a long-distance issue involving 
the IR quantity $1/m$? 
Let me present a conjecture about how this could happen.
This is based on the standard result that $a$ enters into
the EFT through the Symanzik quark-level effective Lagrangian~\cite{Symanzik}.
The IR distance scale thus introduced is $1/(a\Lambda_{\rm QCD}^2)$.
The idea is then that competition between
terms in the EFT proportional to $m$ and $a \Lambda_{\rm QCD}^2$ 
(in fact, it turns out $a^2\Lambda_{\rm QCD}^3$) can cause the condensate 
to ``swing'' from positive to negative values as $m$ is varied.
This is exactly what happens for Wilson fermions~\cite{Creutzaoki,ShSi},
leading to the Aoki phase,
but a similar phenomenon
can also occur with staggered fermions~\cite{LeeSharpe,rSChPT,Aubinaoki}.
There are many details to be worked out, 
so this is a speculation, but such an explanation
would lead to two general conclusions.
First, the width of the dip region would be $\Delta m\sim a^2$
implying that the would-be zero-modes have $\lambda_{1,2}^\stag \sim a^2$
(rather than $\sim a$).
Second, the explanation would exhibit the non-commutativity 
of $m\to0$ and $a\to 0$ limits,
and thus be consistent with the BGSS analysis.
I stress that the lack of commutatitivy is a general phenomenon,
not necessarily related to rooting.
It occurs, for example, in the
flavor-singlet condensate with two-flavors of Wilson fermions
(although this is essentially
impossible to see in simulations because divergent 
short-distance contributions swamp those of a non-perturbative origin).
Clearly it is important to further develop this potential r\schpt\
explanation of the dip.

\bigskip

I now turn briefly to a different, though related, concern raised by 
Creutz~\cite{Creutz06}:
does the $U(1)_\epsilon$ symmetry lead to Ward identities in the rooted
theory which are inconsistent with the proposed continuum limit?
More generally, does the extra, unphysical taste symmetry of $D_\stag$
lead to unwanted relations between correlation functions in the rooted
staggered theory? The answers are no~\cite{BGSS}.
While there are additional symmetries and corresponding Ward identities,
these have no impact on those correlation functions that become
physical in the continuum limit, assuming that this limit is QCD. 

The lack of inconsistency can be seen simply from the properties of
r\schpt. This EFT has an exact $U(1)_\epsilon$ symmetry when $m=0$,
and thus has the corresponding exact Ward identities, but nevertheless,
when $a\to 0$, it has a unitary subsector which is \chpt\ for QCD.

Rather than elaborating on r\schpt, however, I think it is clearer
to use a formal description of the expected continuum theory
with rooted staggered fermions, in which the fermionic weight is\footnote{%
This simplifies the form that arises from the RG analysis
by leaving out the tower of blocked gauge fields.
These play no role, however, in the symmetries that are used 
in the present argument. A less formal discussion can be
developed using the reweighted theories, along the lines
followed in app.~\ref{app:valence}.}
\begin{equation}
{\det}^{1/4}\left[(\widetilde D_\cont\otimes {\bf 1})+ J\right]
=
\det[\widetilde D_\cont] 
\exp\left\{\frac14 \tr\ln[1 + J (\widetilde D_\cont\otimes {\bf 1})^{-1}]\right\}
\,.
\label{eq:BGSS}
\end{equation}
The idea is that $D_\stag$ goes over to the taste 
symmetric form $(\widetilde D_\cont\otimes {\bf 1})$.
The mesonic sources $J$ allow the
construction of correlation functions
(in the same way that the quark masses were used in the \chpt\ 
analysis of sec.~\ref{sec:ChPT}).
The sources here 
have general spin, flavor and taste structure.
The taste structure means that, although the
fermionic weight (when $J=0$) is that of the
desired one-taste theory, the correlation functions
access an extended partially quenched theory
involving the extra tastes.
Exact Ward identities between these correlators can be derived
by doing symmetry transformations on the sources.
These include those noted by Creutz, which follow from
performing $U(1)_\epsilon$ transformations 
[corresponding to spin-taste $(\gamma_5\otimes\xi_5)$],
but also many others because
of the extended $SU(4)$ taste symmetry.
Despite this extended structure, if one uses taste-singlet sources
the fermionic weight reduces to that of QCD with physical sources:
\begin{equation}
J \longrightarrow (\widetilde J\otimes {\bf 1}) 
\qquad\Rightarrow\qquad
{\det}^{1/4}\left[(\widetilde D_\cont\otimes {\bf 1})+ J\right]
\longrightarrow
\det[\widetilde D_\cont + \widetilde J]
\,.
\end{equation}
Thus the physical subsector of the extended PQ theory is
not constrained by the additional Ward identities.
In particular, the precise manner in which the additional Ward identities
are saturated is unimportant for the physical subsector.
For more details, and explicit examples, see Refs.~\cite{BGSS,BGSlat06}.

\bigskip
What is the upshot of this discussion?
Two peculiar properties of rooted staggered fermions have
been identified:
their behavior when one quark mass vanishes and their need
for an extended, PQ continuum limit.
Are these diseases (indicating a wrong theory) or ugly features? 
I think it is now clear that they are ugly
features, {\em as long as one accepts as plausible the
arguments of sec.~\ref{sec:tame} that the continuum limit is correct.} 
Let me repeat this logic, lest it get lost in the details:
the BGSS arguments show that if the continuum limit is correct,
then the peculiar properties are expected.
If, however, one looks just at the required behavior of
the eigenvalues of $D_\stag$, or just at the way in which the
``extra'' Ward identities must be satisfied 
(including contributions from ghosts since the theory is PQ), 
then they may appear implausible.
This is not my reaction,  but it is that of Creutz~\cite{Creutzlat06}.
What I want to stress, however, is that
this is not the right place to hold the argument over rooting---that
belongs in the previous section. The peculiar properties must hold
if the continuum limit is correct, and cannot be used to prove the
rooting trick wrong by contradiction.

\section{Is there a ``valence rooting'' problem?}
\label{sec:valence}

In this section I discuss what I have called
above the ``valence rooting'' issue.
The basic question is how to
account for the extra tastes present in valence propagators,
although I will pose it in a sharper form shortly.
This is an important issue since
extensive calculations with valence staggered fermions are being
done (leading to several of the ``gold-plated'' results).
It is different from the issue which has been my focus up to now,
namely the effects of using a rooted determinant, 
although the two are closely related.
In particular, any resolution of the valence rooting issue
certainly requires that the rooting prescription 
for the determinant leads to the correct continuum limit,
i.e. that the premises of
the arguments of sec.~\ref{sec:tame} are correct.

Let me state the issue in a more pointed way, quoting
a hypothetical skeptic:
\begin{quote}
{\em 
Let me accept that the rooted staggered determinant
can be recast as a single-taste local theory in the
continuum limit, with Dirac operator $\widetilde D_\cont$.
Does this not mean that the actual calculations
are then being done in a mixed-action theory,
since the valence Dirac operator $D_\stag$ differs
from $\widetilde D_\cont$? 
Furthermore, unlike in a standard mixed-action set-up
we do not know the explicit form of $\widetilde D_\cont$,
and so we cannot calculate sea-quark hadron propagators
in order to match valence and sea quark masses, even
in the continuum limit. 
Does this not imply that calculations in the continuum
limit will be partially quenched with uncontrolled
unitary violations, since we
do not know $m_{\rm valence}-m_{\rm sea}$?}
\end{quote}
The answer to both these question is no. 
The essential reason is that applying
the {\em same} blocking to the valence propagators as
to the $D_\stag$ in the determinant leads to
the {\em same} form $(\widetilde D_\cont\otimes{\bf 1})$.
[I am using here $D_\cont$ as a shorthand for
the  $\widetilde D_{n\to\infty}$ used above.]
Since the same operator appears in the valence and sea sectors,
one obtains
$m_{\rm valence}=m_{\rm sea}$ without any tuning.
Thus unitarity violations associated with
$m_{\rm valence}\ne m_{\rm sea}$ will be absent.
It is true, however, that there are additional valence tastes
(because of the ``$\otimes {\bf 1}$'')
and so the continuum theory is partially quenched in this sense.
But, as already discussed in the previous section, 
because $m_{\rm valence}=m_{\rm sea}$ this PQ
theory has a unitary, physical subsector 
(that coupled to taste-singlet mesonic sources) for which
any potential concerns about partial quenching are irrelevant.

This summary, while accurate, glosses over some important
details. Since, as far as I know, there is no thorough discussion
of these details in the literature, I present one
in appendix~\ref{app:valence}.

The valence rooting issue has also been resolved in one 
limited, though important, context, that of
the chiral EFT describing the long-distance physics
of the rooted theory~\cite{CB06}.
If the valence and sea-quark Dirac operators differed one would
perforce be using a mixed action, so the resulting EFT should be of
the mixed-action type~\cite{MAChPT}. 
In particular, there would be differences
between the masses of charged PGBs composed of 
the same flavors of valence and sea quarks.
But this is not the case. 
As discussed in sec.~\ref{sec:ChPT}, we know the EFT
for rooted staggered fermions to be r\schpt,
and this does {\em not} have the mixed-action form~\cite{CB06}.
This is not to say that there are no unphysical effects for $a\ne 0$,
but rather that these effects are not of the form resulting
from the use of a mixed action.

\section{Rooting at finite temperature and density}
\label{sec:GSS}

Rooted staggered fermions have been used extensively
in simulations of the properties of QCD at non-zero temperature ($T$)
and density ($\mu$). Studies at $T>0$ (but $\mu=0$) are in a mature phase,
similar to that of the $T=0$ 
calculations discussed in sec.~\ref{sec:update} above.
They are reviewed here by 
Heller~\cite{Hellerlat06} and Stephanov~\cite{Stephanovlat06},  
Studies at $\mu\ne 0$ are limited by the sign problem and are still exploratory.
They are discussed here in review talks
by Schmidt~\cite{Schmidtlat06}, Splittorff~\cite{Splittorfflat06}, 
and Stephanov~\cite{Stephanovlat06}.
The question I want to address briefly here is whether,
compared to $T=\mu=0$,
additional issues arise when using rooted staggered fermions
at non-zero $T$ and/or $\mu$.

For $T>0$, $\mu=0$, the answer is no.
This is because the PT and RG analyses
discussed in secs.~\ref{sec:PT} and \ref{sec:RG}, respectively,
are done at short distances, and are
unaffected by the boundary conditions on
fields necessary to implement $T>0$.
What could be a problem in practice, however, is
a combination of two factors:
first, that the lattice spacings used at $T>0$ are perforce
larger than those used at $T=0$, implying larger scaling
violations, and, second, that one does not have a \schpt\ form to
fit with in general. 
The chiral EFT can only be used for the relatively uninteresting
regime $T\ll \Lambda_{\rm QCD}\approx T_c$.
These factors make it even more important 
to use improved actions than at $T=0$.

The situation is quite different for $\mu>0$.
It has recently been argued by Golterman, Shamir and
Svetitsky (GSS) that rooted staggered fermions then face
a new problem~\cite{GSS}. A nice summary of their argument
is given in their presentation here~\cite{GSSlat06}.
I give only a brief sketch of this argument and 
then comment on its implications.

The essential point concerns the distribution of
eigenvalues of $D_\stag$. For $\mu=0$,
as already noted in sec.~\ref{sec:diseases},
the eigenvalues lie on a line of the form $m + i\lambda$,
and so do not approach closer than $m$ to the origin.
As $\mu$ increases, however, the eigenvalues spread off
this line into the complex plane.
Once $\mu$ exceeds a critical value the distribution extends
into the region with negative real part, and thus overlaps
the origin. This critical value
is $\mu\approx m_\pi/2$ at $T=0$, but it likely
increases with $T$, although how is presently unclear.

Once eigenvalues overlap the origin the definition
of rooting becomes ambiguous. In particular, the
phase of the determinant becomes ambiguous,
which is problematic because the phase plays an essential role
in $\mu>0$ physics.
To understand this ambiguity, recall
that the eigenvalues are expected to lie in quartets, 
separated by of $O(a)$
[or possibly of $O(a^2)$ for some eigenvalues, as discussed
in sec.~\ref{sec:diseases} and in Ref.~\cite{GSSlat06}]. 
(The quartet structure is not expected to hold if $|\lambda|\sim 1/a$,
but such eigenvalues do not contribute to the uncertainty in the
phase of the determinant.)
If a quartet lies away from the origin, any value lying ``within''
the quartet, such as the average $\bar\lambda$, provides 
a sensible definition of its contribution to the rooted determinant.
All such choices have similar phases.
If, however, the quartet straddles the origin then the phase is ambiguous---
the region within the quartet has, by definition, all possible values
of the phase. 

This ambiguity introduces a systematic error proportional
to the fraction of configurations which have one or more eigenvalue
quartets
straddling the origin. GSS estimate this error to be
$\sim a^3 V^{3/2} \Lambda_{\rm QCD}^9$,
where $V$ is the space-time volume.\footnote{%
This form assumes the splitting in eigenvalue quartets
to be of $O(a)$. The error is parametrically smaller if
the splitting is of $O(a^2)$. This does not, however, effect
the qualitative conclusions drawn in this section.}
The volume appears because the density of eigenvalues increases
with the volume.
This result implies that one must take the continuum limit before
taking the infinite volume limit---another rather ugly feature
of rooted staggered fermions.

How does this impact present calculations at $\mu>0$?
The simulations which are most clearly effected are those 
working directly at ${\rm Re}(\mu)\ne 0$.
The rough estimates of GSS indicate that the problem may
be severe for present simulations,
but this is controversial, and needs further study.
In particular, one needs to look at the distribution
of eigenvalues on the configurations produced
in the actual simulations---do they lie in quartets, and
how frequently do quartets straddle the origin?
The impact on the method
using analytic continuation from imaginary $\mu$
is less clear.
The bottom line is that this is a new issue which must be
added to the travails of finite density studies if one
uses rooted staggered fermions.

\section{Summary}
\label{sec:summ}

In this talk I have tried to provide a rather comprehensive
discussion of the status of rooted staggered fermions, 
concentrating on the core theoretical issues.
There has been a lot of progress in the last year, most
of it ``positive'' for rooting, the main exception being
the new problems that arise for non-zero chemical potential.

Where do things stand? Are rooted staggered fermions good, bad or ugly?
It is clear that we can rule out ``good''---the non-locality 
for $a\ne 0$ ensures a complicated continuum limit. The choice is
between ``bad'' (or ``diseased'') and ``ugly''.
In my opinion, the arguments presented in sec.~\ref{sec:tame}
(using perturbation theory, the renormalization group and \chpt)
make plausible that the correct choice is \dots {\bf ugly}.
Let me repeat what this means---rooted staggered fermions
are unphysical for $a\ne 0$, yet the results go over to those
of a physical, single-taste theory in the continuum limit.
In practice this very likely requires fitting to the complicated,
and one might say ugly, forms of r\schpt.
In this way the unphysical effects are included in the fit,
and in particular in the error estimate.

I have also discussed the important, but secondary, issue
of the impact of treating sea and valence quarks 
differently---crudely speaking, of ``rooting one but not the other.''
I have argued that there is, in fact,
no negative impact, as long as one chooses valence correlators
appropriately and does the appropriate perturbative matching.

If rooted staggered fermions are indeed ``ugly'', 
then the price one pays for
rooting is the need for significantly
more elaborate fitting, and the concomitant increase in
systematic errors. In addition, to obtain the fitting forms
one must carry out the r\schpt\ calculation anew for each quantity,
and this calculation can become
very involved, especially those involving
baryons or complicated operators.
Nevertheless, I stress that, for the mesonic quantities that have
been the main focus to date, these complications have been
successfully handled, and the resulting errors are small.

My conclusion of ``ugly'' rather than ``bad'' clearly rests on an
assessment of the assumptions needed to make the RG argument of
Shamir, and the \chpt\ argument of Bernard.
To me they are plausible, and in some cases very plausible.
Plausibility is, however, in the eye of the beholder.
Others might find the assumptions implausible.
It is thus important to test these assumptions and the
good news is that, to varying extents, this is possible.
The highest priority items are these:
\begin{itemize}
\item
Test the two key assumptions of the RG argument of sec.~\ref{sec:RG}:
\begin{itemize}
\item
The required locality of the blocked Dirac operators and
the effective gauge actions $S^{(k)}_{\rm eff}$ can be
tested using numerical simulations;
\item
The required scaling of the taste non-invariant parts of the blocked
Dirac operators can be tested using perturbation theory and
numerical simulations.
\end{itemize}
\item
Strengthen the foundations of partially quenched \chpt,
which is needed to make the argument of sec.~\ref{sec:ChPT}.
This is a theoretical issue, but is also tested numerically
by PQ simulations with {\em any} type of fermion.
\item
Complete the proof of the renormalizability of unrooted staggered fermions,
upon which the perturbative claims for rooted staggered fermions
are based.
\end{itemize}
Also welcome are further numerical tests of rooting,
in particular the studies of eigenvalues, and further
tests of the validity of r\schpt\ formulae.

Although I have argued that attempts to show the
inconsistency of rooted staggered fermions have failed,
these arguments have clarified the limitations of rooting.
Rooted simulations of QCD are, in practice, restricted to 
having all quark masses positive and none close to vanishing.
The good news is that this allowed regime includes 
physical QCD. The bad news is that rooted studies of theories
other than physical QCD, e.g. one flavor QCD, which are
of interest for a number of reasons~\cite{Degrand}, will
be hampered.

The dispute over rooting has not been, and will likely never be,
completely resolved. This is the nature of working 
in a non-perturbative regime.
The best that one could hope for 
would be to show that rooted staggered fermions have the same continuum limit
as that of one of the uncontroversial fermions,
i.e. that they are on the same footing as other fermions.
While this has not been demonstrated, what I find encouraging
is that, with the RG argument, we have a theoretical framework
in which we can see both how the required unphysical effects are
present and how they vanish in the continuum limit.
In particular, the naive statement that taste symmetry
is restored in the continuum limit, crudely speaking
that $D_\stag$ goes over to $(\widetilde D_\cont\otimes 1)$,
so that the determinant can be fourth-rooted,
can now be backed up with a detailed description of the corrections.

How should a non-lattice physicist react?
The bottom line I think is that there is substantially
more evidence for the validity of rooting now than a year ago.
Irrespective of the rooting debate, however,
any result obtained with one type of fermions alone would
need to be cross-checked with other methods.
So perhaps the best news in the last few years is the major
strides that have been made in simulating with other fermions.
The desired checks will come sooner rather than later.

\section*{Acknowledgments}
I am very grateful to Claude Bernard, Philippe de Forcrand,
Maarten Golterman, Karl Jansen, Tony Kennedy and Yigal Shamir
for many helpful discussions and comments on the manuscript.
This research was supported in part by 
U.S. Department of Energy Grant No. DE-FG02-96ER40956.

\appendix
\section{Perturbation theory with staggered fermions}
\label{app:PT}

In this brief appendix I describe 
some important features of perturbation theory 
for staggered fermions
that are mentioned in the text.

Perturbation theory for (unrooted) staggered fermions was
developed in Ref.~\cite{STW}, and subsequently refined
by many authors.
Because of doubling, the fermion propagator has $2^4$ poles in
the Brillouin zone. The positions of these poles are conveniently
labeled (in lattice units) as $p_\mu = A_\mu \pi$,
where $A_\mu$ is a hypercube vector (all components $0$ or $1$).
Each pole is associated with a reduced Brillouin zone, half as wide
in all directions ($-\pi/2< p'_\mu \le \pi/2$).
A general momentum can then be packaged as $p = p' + A \pi$,
in which $p'$ turns out to be 
the physical momentum and $A$ the spin-taste label.
This construction is similar to the position space decomposition
described in sec.~\ref{sec:SF}, but differs because the
phase associated with a physical momentum here varies within
a spatial hypercube, while that in the position-space construction
does not.

An attractive feature of this ``momentum-space basis'', as it is called
for obvious reasons, is that the propagator is taste symmetric:
\begin{equation}
G^{-1}(q'+ B \pi,p' + A \pi)
= \overline\delta(q'+p')
\left[\sum_\mu i \sin q'_\mu \overline{\overline{(\gamma_\mu\otimes {\bf 1})}}_{BA}
+ m \overline{\overline{({\bf 1}\otimes {\bf 1})}}_{BA}\right]
\,.
\label{eq:SFprop}
\end{equation}
Here $\overline\delta$ is the periodic $\delta-$function
(with period $2\pi$), and the ``double-barred'' spin-taste matrices
are unitarily equivalent to the matrices $(\gamma_B\otimes \xi_C)$
appearing in the position-space basis, eq.~(\ref{eq:SstagQ}).
This packaging was introduced by Ref.~\cite{GS} and
refined in Refs.~\cite{DK,DS}.
The result (\ref{eq:SFprop}) is to be contrasted
to the position-space basis action of eq.~(\ref{eq:SstagQ}),
which has the $O(a)$ taste-breaking term.
There is no contradiction between these two results.
In particular, the $O(a)$ term
in eq.~(\ref{eq:SstagQ}) can be removed by a field definition
and thus does not contribute to on-shell quantities~\cite{Luo1}.
Nevertheless, the difference between the results in the two bases
illustrates why a study of taste-breaking using PT
is most straightforward in the momentum-space basis.

Taste {\em is} broken by vertices, and I want
to show a simple example since this is referred to
in sec.~\ref{sec:PT}. This is the
$\bar q q g$ vertex, which takes the form~\cite{SSbook}
\begin{equation}
V_\mu(\underbrace{q'\!+\!\pi B}_{q},\underbrace{p'\!+\!\pi A}_{\bar q},
\underbrace{k'\!+\!\pi C}_{g})
= -i g\; \overline\delta(p'+q'+k') \cos(q'_\mu + k'_\mu/2)\;
\overline{\overline{(\gamma_{\mu \widetilde C}\otimes \xi_{\widetilde C})}}_{AB}
\,,
\end{equation}
assuming that $C_\mu=0$ ($\mu$ being the gluon polarization direction).
The new hypercube vector  is
$\widetilde C_\nu =_2 \sum_{\rho\ne\nu} C_\rho$, 
the important property of which is that $\widetilde C=0$ iff $C=0$.
All primed momenta (including that of the gluon) lie in the reduced
Brillouin zone (although the result actually holds independent of this).

Taste is broken by the taste matrix $\xi_{\widetilde C}$,
but only if $C\ne0$. This in turn implies that the gluon
has a large, unphysical momenta ($\sim \pi/a$ in physical units),
and its contribution is suppressed by the gluon action.
Gluons with near-physical momenta have $C=0$ and conserve taste.

\section{Resolution of valence rooting issue}
\label{app:valence}

In this appendix I sketch how the ``valence rooting issue'',
which has been stated in sec.~\ref{sec:valence}, can be resolved.
The argument makes essential use of Shamir's RG analysis~\cite{ShamirRG06},
and also draws on work done long ago when using staggered
fermions to calculate electroweak matrix elements
(see, e.g., Refs.~\cite{stagME}). See also the related
discussion in Ref.~\cite{ShamirRG04}.
I assume that rooted staggered fermions lead to a single-taste,
physical continuum limit, and in particular that 
Shamir's RG analysis (discussed in sec.~\ref{sec:RG}) is correct.
In this sense the valence rooting issue is secondary.

As noted in sec.~\ref{sec:valence}, the issue is that
the valence propagators used in practice, $D_\stag^{-1}$,
are not obviously related to the single-taste operator which resides
in the determinant in the continuum limit, $\widetilde D_{n\to\infty}$.
The argument of this appendix attempts to establish this relation.
There are two parts to the argument: first, dealing with the
additional tastes that are present in $D_\stag^{-1}$, and,
second, showing how blocking relates the valence Dirac operators in 
essentially the same way as those appearing in the determinant.

Recall from sec.~\ref{sec:RG} that
the blocked form of $D_\stag$
can be written as a taste invariant piece plus a remainder
which vanishes with $a_f$: $D_n = (\widetilde D_n\otimes {\bf 1}) + O(a_f)$.
The remainder is bounded, allowing the rooted
determinant to be expanded as $\det(D_n)^{1/4}=\det(\widetilde D_n)[1 + O(a_f)]$.
The propagator can be expanded similarly,\footnote{%
Strictly speaking, this holds only in the ensemble of configurations
produced with the gluonic and fermionic weights. 
In particular, as noted in sec.~\protect\ref{sec:RG}
and app.~\protect\ref{app:PT}, the damping
provided by the gluon action is necessary to obtain the smallness of
the correction term.}
\begin{equation}
D_n^{-1}(x,y) = (\widetilde D_n\otimes {\bf 1})^{-1}(x,y) [1 + O(a_f)]
\,,
\label{eq:DnvstildeDn}
\end{equation}
with the taste non-invariant correction vanishing as $a_f\to 0$ for fixed 
physical quark mass and fixed physical distance $x-y$.
In the following, I take $n$ large enough (i.e. take $a_f$ small enough relative to 
the fixed coarse lattice spacing $a_c$) that the
$O(a_f)$ corrections are negligible.
It is convenient to make use of Shamir's reweighted theories, in which
one changes the fermionic weight from $\det(D_n)^{1/4}$ to
$\det(\widetilde D_n)$. I extend this by using propagators
$(\widetilde D_n\otimes {\bf 1})^{-1}$ in place of $D_n^{-1}$.
This allows one to work with an exactly taste-invariant theory at finite $n$,
and yet obtain results differing only by corrections of $O(a_f)$
from those in the original rooted theory with propagators $D_n^{-1}$.

The reweighted theory with propagators $(\widetilde D_n\otimes {\bf 1})^{-1}$
has the correct number of sea quarks and thus
differs from conventional fermions only by the presence
of extra tastes in the propagators.
This does mean, however, that it is partially quenched. 
The ``fermionic'' part of the action can be written\footnote{%
This formulation is similar to, but more general than, that
used in sec.~\ref{sec:diseases}---see eq.~(\protect\ref{eq:BGSS}).
Here the sources couple to quark fields rather than to mesons,
allowing one to create the more general correlators of the type
used to calculate matrix elements and baryons.}
\begin{equation}
{\cal L}_F 
=
\sum_{i=1}^{4} 
\left\{\widetilde{\bar\psi_i} \widetilde D_n \widetilde \psi_i
+ \bar\eta_i \widetilde \psi_i + \widetilde{\bar\psi_i} \eta_i \right\}
+ \sum_{j=1}^{3}
\phi_j^\dagger \widetilde D_n \phi_j
\,, \label{eq:PQreweightedS}
\end{equation}
where the $\widetilde\psi_i$ are four single-taste fermions, 
while the $\phi_j$ are ghosts (commuting spin-$1/2$ fields) 
corresponding to {\em three} of these fermions. 
Having one more fermion than ghost leads to the desired
determinant. The sources $\eta$ and $\bar\eta$ allow propagators to be generated
in the standard way, and since all four tastes are ``sourced''
the general propagator is the full $(\widetilde D_n\otimes {\bf 1})^{-1}$.
I stress that the PQ set-up is needed because
$D_\stag^{-1}$ is related to this four-taste propagator,
and not the single-taste $\widetilde D_n^{-1}$.

It should also be kept in mind that all levels of gauge
field (from fine to coarse) are still present. The
theory can be made to look more conventional
by integrating out all but the coarse lattice fields.
This gives rise to a multitude of multifermion interactions,
but, as noted in sec.~\ref{sec:RG}, does not change the
universality class. It also does not effect the source terms
in the action (\ref{eq:PQreweightedS}), which can still be used
in the standard way to generate correlation functions. 
Nothing essential is gained by integrating out the gauge fields,
however, and one looses the simplicity of the fermion action,
so I choose not to do this integration in the following.

I can now address the issue of the extra valence tastes.
How can one calculate the single-taste quantities of interest?
One approach is straightforward: simply construct
correlators using sources only of a single taste,
say $\eta_4$ and $\bar\eta_4$.
In this way, one can calculate any correlation
function of interest (meson, baryon, matrix elements, \dots)
exactly as for uncontroversial fermions.
This is possible because taste is unbroken in the reweighted theory.
In effect, the procedure replaces $(\widetilde D_n\otimes {\bf 1})^{-1}$
with $\widetilde D_n^{-1}$.
For example, the charged pion correlator would simply be
\begin{equation}
C_\pi^{1-{\rm taste}}(x,y) = 
\langle {\rm Tr}[\gamma_5 
\widetilde D_n^{-1}(x,y) \gamma_5
\widetilde D_n^{-1}(y,x)]\rangle_{n{\rm 'th\ reweighted}}
\,,
\label{eq:Cpiwewant}
\end{equation}
where quark masses and color indices are implicit,
and $x$ and $y$ are points on the coarse lattice.
Note that the determinants arising from integrating out the other
three tastes of fermions and ghosts cancel,
so one is working in
a physical subsector of the partially quenched theory.

In practice, however, the calculations that are done
do not correspond to using the single-taste approach in the
reweighted theory.
Instead, all tastes are used, but in such a way that
the resulting correlator equals (identically) 
that of the single-taste construction.
This equality can be checked by hand on a case-by-case basis---it involves
a purely group-theoretical exercise (really just counting)
since the dynamics is taste-invariant.\footnote{%
Similar ``tricks'' have also been used recently in the
applications of twisted-mass fermions~\cite{FRPQ}, although
in that case one does not face the issue of rooting.}
The flexibility of using all tastes allows one to project external operators
onto representations of the staggered fermion symmetry group,
thus disentangling different contributions to correlators.
For example, for mesonic quantities it is often advantageous
to use taste $\xi_5$, which couples to the lattice Goldstone pion.
In the example of the charged pion propagator one then has
\begin{equation}
C_\pi^{1-{\rm taste}}(x,y) = \frac14
\langle {\rm Tr}[(\gamma_5\otimes{\xi_5}) 
(\widetilde D_n\otimes {\bf 1})^{-1}(x,y) (\gamma_5\otimes {\xi_5})
(\widetilde D_n\otimes {\bf 1})^{-1}(y,x)]\rangle_{n{\rm 'th\ reweighted}}
\,,
\end{equation}
The factor of $1/4$ cancels the sum over tastes.
This approach of averaging over tastes has long been used, e.g. in setting up
the calculation weak matrix elements with staggered
fermions~\cite{stagME}. 
They are the analog of the factors of $1/4$ introduced
in sea-quark loops by rooting.

It should be stressed that the use of reweighted theories allows
the flexibility of using different taste projections to be put on
a firm footing. Previously (and I am guilty of this) one worked out
the counting factors assuming that taste was restored, not taking
into account the fact that taste is actually broken for $a\ne 0$.

The discussion so far has shown how one can deal with
the extra valence tastes. The end result is that the desired
correlators can be built out of propagators 
$(\widetilde D_n\otimes {\bf 1})^{-1}$, with appropriate taste projections
and normalization factors added by hand.
One then uses the result (\ref{eq:DnvstildeDn}) above to replace these propagators
with $D_n^{-1}$, and similarly replaces the reweighted determinant by
the original rooted determinant,
knowing that one makes only an error of $O(a_f)$ when doing so. 
Thus, for the example of the charged pion correlator one has
\begin{eqnarray}
C_\pi^{1-{\rm taste}}(x,y) &=& C_\pi^\stag(x,y) + O(a_f) \,,
\\
C_\pi^\stag(x,y) &=& \frac14
\langle {\rm Tr}[(\gamma_5\otimes{\xi_5}) D_n^{-1}(x,y) (\gamma_5\otimes {\xi_5})
                  D_n^{-1}(y,x)]\rangle_{\rm coarse}
\,.
\label{eq:blockedval}
\end{eqnarray}
The expectation value (labeled ``coarse'')
involves  an integration over all levels of gauge field,
weighted by the original gauge action and blocking kernels 
and by $\det(D_n)^{1/4}$.
Just to be completely clear, the message of eq.~(\ref{eq:blockedval})
is that the (theoretically uncontroversial) correlator
we want to calculate, $C_\pi^{1-{\rm taste}}$,
is related (up to a controlled error) to
a rooted staggered correlator, $C_\pi^\stag$.

Now I can return to the second issue, namely that we  actually calculate using
$D_\stag^{-1}$ rather than $D_n^{-1}$, so that $C_\pi^\stag$ is not yet of
the form we actually calculate.
The key point is that, by ``undoing'' the fermionic blocking,
one can rewrite the blocked propagator in terms of the original fine-lattice
propagator. This is done by repeated use of the result
\begin{equation}
D_k^{-1} = \alpha_k^{-1} + Q^{(k)} D_{k-1}^{-1} Q^{(k)\dagger}\,,
\end{equation}
and leads to the schematic form
\begin{equation}
D_n^{-1}(x,y) = c\; \delta_{x,y} + 
\sum_{x',y'} {\cal Q}(x,x') D_\stag^{-1}(x',y') {\cal Q}^\dagger(y',y)
\,.
\label{eq:DnvsDstag}
\end{equation}
Here $x$ and $y$ are coarse lattice points, while
$x'$ and $y'$ lie on the fine lattice.
The constant $c$ can be expressed in terms of the $\alpha_j$'s.
${\cal Q}$ is the product of the fermion blocking kernels,
and has a range $\sim a_c$. It depends on the gauge fields on the
fine lattice and all the blocked lattices
(except the ``last'' one, $V_n$, although this will not
matter here). 
In words, eq.~(\ref{eq:DnvsDstag}) says that the blocked propagator
is obtained from the original fine-lattice propagator by smearing
over a range $a_c$ on both ends, and adding a contact term.

We next insert the result (\ref{eq:DnvsDstag}) into the
correlator we are calculating, e.g. eq.~(\ref{eq:blockedval}),
and integrate out all but the fine-lattice gauge fields. 
This brings us back to the original fine lattice action 
(that which is simulated), 
including the rooted determinant of $D_\stag$.
The integration also changes the combined kernel ${\cal Q}$,
which depends on the blocked gauge fields.
Consider the correlator (\ref{eq:blockedval}), but
with $|x-y|\gg a_c$, so that the contact term
in $D_n^{-1}$ does not contribute, and the kernels ${\cal Q}$
at the two ends of $D_n^{-1}$ do not overlap. 
Then the gluon integration acts separately on the two ends,
and changes the product of kernels into a smeared source:
\begin{equation}
{\cal Q}^\dagger(x'',x)(\gamma_5\otimes {\xi_5}) {\cal Q}(x,x')
\longrightarrow
{\cal S}_5(x'',x,x')
\,.
\end{equation}
Here ${\cal S}_5$ is a sum 
(weighted by phases) over products of links
running from $x''$ to $x$ and then to $x'$, possibly
multiplied by nearby closed Wilson loops.
It is no doubt horribly complicated, but we do not need
its explicit form. Inserting this result into (\ref{eq:blockedval})
we end up with
\begin{equation}
C^\stag_\pi(x,y) =
\langle {\rm tr}[{\cal S}_5(x'',x,x') D_\stag^{-1}(x',y')
{\cal S}_{5}^\dagger(y',y,y'') D_\stag^{-1}(y'',x'')]\rangle_{\rm fine}
\,,\qquad |x-y|\gg a_c \,.
\label{eq:CpiDstag}
\end{equation}
Thus we learn that the correlator we want to calculate,
eq.~(\ref{eq:blockedval}), is, at long distances,
equal to a correlator of the form that is actually calculated, i.e.
involving $D_\stag^{-1}$.
This is an example of the ``pull-back mapping'' used extensively by Shamir.

We are not quite done. The result (\ref{eq:CpiDstag}) involves
the mesonic sources ${\cal S}_5$ rather than the sources
used in practice. The former are smeared (size $\sim a_c$), 
complicated and not known explicitly, while the latter are
typically ultralocal (size $\sim a_f$).
In a physical theory this would not matter, at least for
determining the states that are created.
One need only ensure that the sources used in practice have the
same transformation properties under the lattice symmetry
group as the ${\cal S}_5$, properties which are determined by
the operator that is inserted in the original coarse-lattice
correlator [here $(\gamma_5\otimes {\xi_5})$].
Using a spectral decomposition, this is sufficient to ensure
that the actual correlator couples to the same states as contribute to
(\ref{eq:CpiDstag}).

This argument does not go through as stated with rooted
staggered fermions---it is the theory after, not before, blocking which
has been argued to be physical (up to $O(a_f)$ corrections).
To make progress, I think one must rely on perturbation theory.
Perturbation theory for rooted staggered fermions has been
briefly discussed in secs.~\ref{sec:BGS} and \ref{sec:PT}.
It involves the replica trick for fermion loops, but keeps
the full $D_\stag^{-1}$ on valence lines. 
It is thus partially quenched,
but in a controlled way since valence and sea propagators
are the same (including the mass terms). Furthermore, by an extension of
the argument in sec.~\ref{sec:PT}, the results of
this ``rooted'' PT become equal [up to errors of $O(a_c)$]
to those in a continuum taste-invariant PQ
theory, i.e. one with the correct number of sea quarks
but with four times as many valence quarks.
Using this perturbation theory, one can,
in principle, relate the correlators
involving ultralocal and smeared operators.
As long as the quarks and gluons created
by these operators have momenta satisfying $|p_i| \ll 1/a_c$,
then they will ``see'' the smeared operator as local. This,
plus gauge invariance and lattice symmetries,
implies that smeared and local operators are related by 
$Z$-factors independent of the external state.
This holds up to corrections suppressed by $a_c |p|$,
which, since the non-perturbative momenta of interest have
$p\sim \Lambda_{\rm QCD}$, requires $a_c \Lambda_{\rm QCD} \ll 1$.

The upshot is that the non-perturbative
long-distance part of the correlators which are
actually calculated is proportional to that of the
correlators involving smeared sources,
such as (\ref{eq:CpiDstag}), with the proportionality
constant calculable in principle using PT.
The latter correlators are themselves equal, up to $O(a_f)$ corrections,
to correlators in a physical single-taste theory,
such as (\ref{eq:Cpiwewant}).
Thus the correlators actually calculated have no ``valence rooting'' problem
in the continuum limit.\footnote{%
This assumes, of course, that the mass
in the valence propagators is the same as that in the corresponding
sea-quark rooted determinant. One can extend this set-up to partially
quenched rooted staggered fermions, and use PQ r\schpt\ to control
the unitarity violations.}

Missing from this discussion is the issue of whether one
can make use of the normalization of the correlators actually calculated.
This is important because these normalizations are
needed to determine matrix elements such as $f_K$ and $B_K$.
In fact one can use the normalization.
The two parts of the argument above---first, matching correlators
to a theory with extra tastes and, second, using PT to match operators at
the coarse and fine scales---can be combined into
the single step of matching the original fine lattice correlator to 
one in a PQ continuum theory, using PT.
What makes this possible is that taste is unbroken in rooted staggered
perturbation theory (up to scaling violations), so the intermediate
step of passing to the reweighted theory at scale $a_c$ is not needed.
The direct ``fine-lattice to PQ continuum'' matching is what is used
in practice in (rooted) staggered calculations of matrix elements.

\end{document}